\documentclass[11pt]{article}
\usepackage{graphicx}
\usepackage{cite}
\usepackage{amssymb}
\usepackage{hep}
\usepackage{feynmp}
\usepackage{multicol}
\usepackage{simplewick}
\usepackage{appendix}
\usepackage{subfigure}
\usepackage{color}
\newcommand{\eins}{\mbox{$1 \hspace{-1.0mm} {\bf l}$}}

\numberwithin{equation}{section}

\begin{document}

\title{\bf Polarized hadron pair production from\\
electron-positron annihilation} 

\author{D.~Pitonyak$^{1}$,
 M.~Schlegel$^{2}$, A.~Metz$^{3}$
 \\[0.3cm]
{\normalsize\it $^1$RIKEN BNL Research Center, Brookhaven National Laboratory, Upton, NY 11973, USA} \\[0.15cm]
{\normalsize\it $^2$Institute for Theoretical Physics, T\"ubingen University,} \\ 
{\normalsize\it Auf der Morgenstelle 14, D-72076 T\"ubingen, Germany}\\[0.15cm] 
{\normalsize\it $^3$Department of Physics, Barton Hall,
  Temple University, Philadelphia, PA 19122, USA }}

\date{\today}
\maketitle

\begin{abstract}
\noindent We study the production of two almost back-to-back hadrons from the annihilation of an electron and a positron, allowing for the polarization of all particles involved.  In particular, we conduct a general (model-independent) structure function decomposition of the cross section for the case $e^+e^-\! \to \gamma^* \to h_ah_bX$.  Moreover, using the parton model we calculate the relevant structure functions in terms of twist-2 transverse momentum dependent (TMD) fragmentation functions (FFs).  We also give results for the situation $e^+e^-\! \to Z^* \!\to h_ah_bX$ (including $\gamma$-$Z$ interference) within this model.  This is the first time a complete framework has been presented for the examination of TMD FFs within $e^+e^-\!\to h_ah_bX$.  We also specify certain parts of our analysis that hold for the triple-polarized semi-inclusive deep-inelastic scattering process and for di-hadron fragmentation.  Furthermore, we give an explicit prescription of how our work can be translated to the Drell-Yan reaction, which provides for the first time full results for double-polarized Drell-Yan that include electroweak effects. We further discuss the relevance of our $e^+e^-\!\to h_ah_bX$ results for future experiments at $e^+e^-$ machines.
\end{abstract}

\allowdisplaybreaks

%
%
%
\section{Introduction}
\label{s:intro}

\noindent The cross section for electron-positron annihilation into hadrons where one does not detect a specific hadron in the final state was an early test for perturbative QCD (see, e.g., \cite{Bethke:1993aj}).  However, such analyses are unable to access the internal (long distance) structure of hadrons.  On the other hand, if one or more hadrons are identified in the final state, then these inner-workings can be probed.  In particular, one can strictly study fragmentation functions (FFs), which embody the process of a parton forming a hadron, and they contain important information about the strong interaction in the non-perturbative regime.  Both the Belle Collaboration at KEK in Japan and the BABAR Collaboration at SLAC in the~US have $e^+e^-$ machines and have performed measurements of a certain azimuthal asymmetry that occurs in electron-positron annihilations when two charged unpolarized almost back-to-back pions are detected in the final state, i.e., $e^+e^-\!\rightarrow \pi^+\pi^-\,X$ \cite{Abe:2005zx, TheBABAR:2013yha}.  As shown in Ref.$\!$\cite{Boer:1997mf}, this asymmetry gives access to the Collins function $H_1^\perp$ \cite{Collins:1992kk}, which describes the fragmentation of a transversely polarized quark into an unpolarized hadron.  (See also \cite{Boer:2008fr} for a comprehensive review of this process in the context of extracting the Collins function.)  Along with an asymmetry involving the Collins function and the transversity $h_1$ that has been determined in semi-inclusive deep-inelastic scattering (SIDIS) \cite{Airapetian:2004tw, Alexakhin:2005iw, Qian:2011py}, extractions of both functions have been performed \cite{Vogelsang:2005cs, Efremov:2006qm, Anselmino:2007fs}.  The one especially of the transversity parton distribution function (PDF) represents a milestone in transverse spin physics.  Numerous other asymmetries also exist in the $e^+e^-\!\rightarrow h_a h_bX$ cross section that involve the polarization of one or both detected hadrons.  Such asymmetries could be readily studied in a situation involving hyperons, e.g., $e^+e^-\!\rightarrow \pi\Lambda\,X$ or $e^+e^-\!\rightarrow \Lambda\bar{\Lambda}\,X$, since through its weak decay the spin of the hyperon can be reconstructed.  Little information exists on polarized FFs --- see, e.g.,\cite{Anselmino:2000vs} and references therein.  Therefore, measurements detecting transversely and longitudinally polarized hadrons could provide valuable insight into the fragmentation process.  

In this paper we focus on the angular distribution of the electron-positron annihilation cross section for the production of almost back-to-back polarized hadron pairs.  This process has been analyzed before for the situation $e^+e^- \!\to \gamma^* \to h_ah_bX$ \cite{Boer:1997mf, Boer:2008fr} as well as $e^+e^-\! \to Z^* \to h_ah_bX$ \cite{Boer:2008fr, Boer:1997qn}, and we will extend upon these previous works.  (Note that the case where only one hadron is detected has also been studied recently \cite{Wei:2013csa}.) To be specific, in part of the work \cite{Boer:1997mf}, the authors, using a diagrammatic approach, evaluate the hadronic tensor in a specific frame up to twist-3 accuracy and compute the fully differential cross section to twist-2 accuracy.  We first discuss in Sect.$\,$\ref{s:DecompHad} how constraints on the hadronic tensor that enters into $e^+e^- \!\to \gamma^* \to h_ah_bX$ enable us to determine its general form, which allows the fully differential cross section to be written valid to any twist and in any frame.  Note that this general form of the hadronic tensor can be readily used in triple-polarized SIDIS (i.e., beam, target, detected hadron all polarized) and for di-hadron fragmentation.  We then write down the cross section for $e^+e^- \!\to \gamma^* \to h_ah_bX$ in Sect.$\,$\ref{s:CSnRef} after a discussion about reference frames.  We next in Sect.$\,$\ref{s:SFT2} use the parton model to calculate the structure functions to twist-2 accuracy for both the reaction involving $\gamma^*$ and the one involving $Z^*$.  In particular, for the latter process we allow for both hadrons to be polarized and also include $\gamma$-$Z$ interference terms.  Note that Ref.$\!$\cite{Boer:1997qn} only considered $Z$-$Z$ terms with one hadron polarized\footnote{In principle, one can also obtain the $\gamma$-$Z$ terms for single hadron polarization from the work in \cite{Boer:1999mm} on Drell-Yan.} $\!$while Ref.$\!$\cite{Boer:2008fr} included $\gamma$-$Z$ interference but only for unpolarized hadrons.  In none of the works \cite{Boer:2008fr, Boer:1997qn, Boer:1999mm} were polarized leptons and/or double hadron polarization included in the electroweak case.  This is the first time a complete framework has been presented for the examination of TMD FFs within $e^+e^-\!\to h_ah_bX$.  We mention that the electroweak process especially would be relevant for a proposed International Linear Collider (ILC) \cite{Baer:2013cma}.  Finally, in Sect.$\,$\ref{s:DandC} we give an explicit prescription of how our work can be translated to the Drell-Yan reaction.  This again is the first time full results are available for double-polarized Drell-Yan that include electroweak effects.  Such experiments could be performed at the Relativistic Heavy Ion Collider (RHIC).  We also comment on some advantages of including lepton polarization in $e^+e^-\!\to h_ah_bX$. Some of these involve new structure functions that have not appeared in the literature before, and, in particular, allow one to test the TMD evolution formalism.  We end with some additional concluding remarks.

%
%
%
\section{Decomposition of the hadronic tensor}
\label{s:DecompHad} 

\begin{figure}[t]
\begin{center}
\includegraphics[width=11cm]{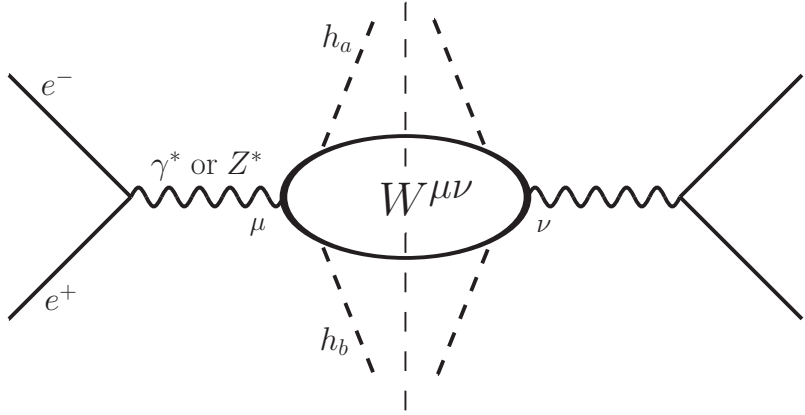}
\caption[]{Cross section for $e^+e^-\!\rightarrow h_a\, h_b\, X$ in terms of its leptonic and hadronic parts.  The leptonic piece contains the (squared) electron-positron interaction, and the hadronic factor contains the (squared) decay of the virtual boson into the two detected hadrons and other (unobserved) particles.  The former can be calculated perturbatively, while the latter is non-perturbative and can be parameterized.  See text for details.}
 \label{f:epluseminusFact}
\end{center}
\end{figure}

\noindent To be definitive, we consider the process
\begin{equation}
e^+(l^{\prime},\lambda') + e^-(l,\lambda) \rightarrow (\gamma^*(q)\;{\rm or}\; Z^*(q)) \rightarrow h_a(P_a,S_a) + h_b(P_b,S_b) + X\,, 
\end{equation}
where the momenta and polarizations of the particles are indicated.  The momentum $P_a$ and spin $S_a$ satisfy $P_a^2 = M_a^2$, $S_a^2 = -1$, $P_a\cdot S_a = 0$ and likewise for $P_b$, $S_b$.  The helicities $\lambda,\,\lambda'$ of the leptons satisfy $\lambda=\lambda' \equiv \lambda_e$.  The differential cross section can be written as the contraction of a leptonic tensor $L^{\mu\nu}$ with a hadronic tensor $W^{\mu\nu}$ \cite{Boer:1997mf, Boer:1997qn, Boer:2008fr} (as we show in Fig.$\,$\ref{f:epluseminusFact}):
\begin{equation}
4\frac{P_a^0P_b^0\,d\sigma} {d^3\vec{P}_a d^3\vec{P}_b} = \frac{2\alpha_{em}^2} {q^2}\left(L_{\mu\nu}W^{\mu\nu}\right)_{\gamma\gamma} + \frac{M_Z ^4 G_F^2}{64\pi^2 q^2}\left(L_{\mu\nu}W^{\mu\nu}\right)_{ZZ} +\frac{\alpha_{em}\sqrt{2}M_Z^2G_F}{8\pi q^2} (\left(L_{\mu\nu}W^{\mu\nu}\right)_{\gamma Z} + h.c.)\,, 
\label{e:epluseminusCross}
\end{equation}
where (including an average over $\lambda^\prime$)
\begin{align}
L^{\mu\nu}_{\gamma\gamma}(l,l^{\prime};\lambda_e) & = \frac{1}{q^4}\big(l^{\prime{\mu}}l^\nu + l^{\prime{\nu}}l^{\mu} - q^2g^{\mu\nu}/2 - i\lambda_e\epsilon^{\mu\nu\rho\sigma}l_\rho l^{\prime}_\sigma\,\big), \label{e:leptens2} \\
L^{\mu\nu}_{ZZ}(l,l^{\prime};\lambda_e) & = \, \frac{1+a_Z^2+2\lambda_e a_Z}{(q^2-M_Z^2)^2+\Gamma_Z^2M_Z^2}\big( l^{\prime{\mu}}l^\nu + l^{\prime{\nu}}l^{\mu} - q^2g^{\mu\nu}/2 - i\lambda_e\epsilon^{\mu\nu\rho\sigma}l_\rho l^{\prime}_\sigma\,\big), \label{e:leptens_ZZ} \\
L^{\mu\nu}_{\gamma Z}(l,l^{\prime};\lambda_e) & = \,-\frac{\lambda_e + a_Z}{q^2(q^2-M_Z^2-i\Gamma_ZM_Z)}\big( l^{\prime{\mu}}l^\nu + l^{\prime{\nu}}l^{\mu} - q^2g^{\mu\nu}/2 - i\lambda_e\epsilon^{\mu\nu\rho\sigma}l_\rho l^{\prime}_\sigma\,\big),\label{e:leptens_phZ}
\end{align}
and
\begin{equation}
W^{\mu\nu}_{\alpha \beta}(q;P_a,S_a;P_b,S_b) = \frac{1} {(2\pi)^4}\sum_{X}\hspace{-0.55cm}\,\int\,(2\pi)^4\,\delta^{(4)}(q-P_X-P_a-P_b)H^{\mu\nu}_{\alpha \beta}(P_X;P_a,S_a;P_b,S_b)\,, \label{e:hadtens} 
\end{equation} 
with
\begin{equation}
H^{\mu\nu}_{\alpha \beta}(P_X;P_a,S_a;P_b,S_b) = \langle 0|J^\nu_{\beta}(0)|P_X;P_a,S_a;P_b,S_b\rangle\langle P_X; P_a,S_a; P_b,S_b|J^\mu_{\alpha}(0)|0\rangle\,, \label{e:Htens} 
\end{equation}
\hspace{-0.15cm}where $\alpha$ and $\beta$ indicate the gauge boson species, i.e., $\alpha, \beta \in \{\gamma,Z\}$.  The factor $J^\mu_{\alpha}$ in Eq.$\,$(\ref{e:Htens}) is the current operator associated with the gauge boson $\alpha$.  The fine structure constant is given by $\alpha_{em} = e^2/4\pi\approx 1/137$, and the weak coupling is related to the Fermi constant $G_F$ via $\alpha_{ew}=\sqrt{2}\,G_{F}M_{W}^{2}/\pi\approx0.034$, with the $W$-boson mass $M_{W}\approx80.389\,\mathrm{GeV}$. On the lepton side we have $a_{Z}=-1+4\sin^{2}\theta_{W}$, with the Weinberg angle $\theta_W$ given by $\sin^{2}\theta_{W}\approx0.231$, $\cos\theta_{W}\approx M_{W}/M_{Z}$, where the mass of the $Z$-boson is $M_{Z}\approx91.188\,\mathrm{GeV}$. We also implement a $Z$-boson decay width $\Gamma_{Z}\approx2.495\,\mathrm{GeV}$ as an imaginary part into the $Z$ propagator.  

We will restrict our general (model-independent) discussion of the hadronic tensor (and the cross section in Sect.$\,$\ref{s:CSnRef})  to the pure electromagnetic case (i.e., only the $\gamma\gamma$ term in Eq.$\,$(\ref{e:epluseminusCross})).  Note that this analysis of the hadronic tensor also holds for triple-polarized SIDIS.  Moreover, the result is not limited to the case where the hadrons are almost back-to-back but, in particular, also applies to di-hadron fragmentation.  Di-hadron FFs have gained attention over the years because, for example, they allow one to access the transversity function within collinear factorization \cite{Collins:1993kq, Jaffe:1997hf, Radici:2001na, Boer:2003ya}.  Such an extraction of the transversity function has been performed recently \cite{Bacchetta:2012ty}.  However, since more convenient reference frames are chosen for di-hadron fragmentation \cite{Bianconi:1999cd, Courtoy:2012ry} than the ones we present in Sect.$\,$\ref{s:CSnRef} (which are more suitable for the back-to-back case), the cross section given there cannot be directly taken over for di-hadron fragmentation.  Likewise, for the triple-polarized SIDIS process one must work, for example, in the target rest frame \cite{Bacchetta:2006tn} in order to obtain a useful result for the cross section.  

The hadronic tensor in (\ref{e:hadtens}) (where now we have $\alpha=\beta=\gamma$) encodes the non-perturbative piece of the reaction and remains an unknown in the process.  Nevertheless, it must satisfy certain constraints, namely, electromagnetic ({\it em}) gauge invariance, parity, and Hermiticity.  These restrictions are quantitatively given by
\begin{align}
q_\mu W^{\mu\nu}(P_a,S_a;P_b,S_b;q) &= q_\nu W^{\mu\nu}(P_a,S_a;P_b,S_b;q) = 0 \;{\rm ({\it em}\, gauge \,invariance)}\,, \\[0.1cm]
W^{\mu\nu}(P_a,S_a;P_b,S_b;q) &= W^{\mu\nu}(\bar{P}_a,-\bar{S}_a;\bar{P}_b,-\bar{S}_b;q) \;{\rm (parity)}\,,\\[0.1cm]
W^{\mu\nu}(P_a,S_a;P_b,S_b;q) &= [W^{\nu\mu}(P_a,S_a;P_b,S_b;q)]^* \;{\rm (Hermiticity)}\,.
\end{align}
Note that $\bar{a}^\mu = a_\mu$ for a generic vector $a$, and we have dropped the $\gamma\gamma$ subscript on the hadronic tensor for brevity.  From these conditions we can break down the hadronic tensor into basis tensors multiplied by structure functions.  Such a decomposition was conducted previously for the Drell-Yan process (and the cross section at twist-2 was calculated) \cite{Arnold:2008kf} (see also \cite{Ralston:1979ys, Tangerman:1994eh, Pire:1983tv, Boer:1997nt, Boglione:2011zw}); we follow a similar procedure here.  However, one additional complication in our case is the inclusion of lepton polarization.  This directly leads to the antisymmetric piece of the leptonic tensor (cf.~Eq.$\,$(\ref{e:leptens2})).  Thus, not only does the symmetric part $W^{S}$ of the hadronic tensor contribute (like in Drell-Yan) but also the antisymmetric term $W^{A}$.  

The approach to writing the hadronic tensor in terms of basis tensors follows along the same lines for both the symmetric and antisymmetric parts.  Here we will restrict ourselves to the antisymmetric piece since the decomposition of the symmetric term was already examined in Ref.$\!$\cite{Arnold:2008kf}.  We will look separately at the situations where the hadrons both are unpolarized, only one ($h_a$ or $h_b$) is polarized, or both are polarized.  First, for the case where both hadrons are unpolarized, we have three available vectors on which the hadronic tensor can depend:~$P_a^\mu$, $P_b^\mu$, $q^\mu$.  This leads to the following basis consistent with the parity constraint:
\begin{align}
h_{U,1}^{A,\mu\nu} &= P_a^\mu P_b^\nu - P_a^\nu P_b^\mu, \label{e:hU1} \\
h_{U,2}^{A,\mu\nu} &= P_a^\mu q^\nu - P_a^\nu q^\mu, \label{e:hU2} \\
h_{U,3}^{A,\mu\nu} &= P_b^\mu q^\nu - P_b^\nu q^\mu, \label{e:hU3}
\end{align}
where the superscript $A$ indicates that these are antisymmetric tensors and the subscript $U$ indicates that both hadrons are unpolarized.  We now impose $em$ gauge invariance through a method introduced in Ref.$\!$\cite{Bardeen:1969aw} by using a projection tensor defined as
\begin{equation}
\mathcal{P}^{\mu\nu} = g^{\mu\nu} - \frac{q^\mu q^\nu} {q^2}\,.
\end{equation} 
We allow this operator to act on the tensors in Eqs.$\,$(\ref{e:hU1})--(\ref{e:hU3}) as follows:
\begin{equation}
h_{U,i}^{A,\mu\nu} \rightarrow \mathcal{P}^\mu_{\,\,\,\,\rho} \,h_{U,i}^{A,\rho\sigma} \,\mathcal{P}_{\!\sigma}^{\,\,\,\nu}. 
\end{equation}
Notice that $q_\mu \mathcal{P}^{\mu\nu} = \mathcal{P}^{\mu\nu} q_\nu = 0$ so that only (\ref{e:hU1}) survives this projection.  We now have
\begin{equation}
t_{U,1}^{A,\mu\nu} = \tilde{P}_a^\mu \tilde{P}_b^\nu - \tilde{P}_a^\nu \tilde{P}_b^\mu, \label{e:tU1}
\end{equation}
where $\tilde{a}^\mu \equiv a^\mu - q^\mu\,q\cdot a/q^2$, and we use the symbol $t$ for the tensors that form our final basis.  Thus, we can write
\begin{equation}
W_U^{A,\mu\nu} = i V^A_{U,1} \, t_{U,1}^{A,\mu\nu}, \label{e:UAhadtens}
\end{equation}
where $V^A_{U,1}$ is a real valued function (structure function) depending on scalar variables of the reaction.  The factor of $i$ shows up in Eq.$\,$(\ref{e:UAhadtens}) because the Hermiticity constraint requires the coefficients of the antisymmetric basis tensors to be pure imaginary.  For the symmetric part of the hadronic tensor, we simply state the result from Ref.$\!$\cite{Arnold:2008kf} obtained from the same procedure outlined above:
\begin{equation}
W_U^{S,\mu\nu} = \sum_{i=1}^4 V^S_{U,i} \, t_{U,i}^{S,\mu\nu}, \label{e:UShadtens}
\end{equation}
where
\begin{align}
t_{U,1}^{S,\mu\nu} &= g^{\mu\nu} - \frac{q^\mu q^\nu} {q^2}\,,\\
t_{U,2}^{S,\mu\nu} &= \tilde{P}_a^\mu \tilde{P}_a^\nu\,, \\
t_{U,3}^{S,\mu\nu} &= \tilde{P}_b^\mu \tilde{P}_b^\nu\,,\\
t_{U,4}^{S,\mu\nu} &= \tilde{P}_a^\mu \tilde{P}_b^\nu + \tilde{P}_a^\nu \tilde{P}_b^\mu\,.
\end{align}
Note again $V^S_{U,i}$ is real valued and depends on scalar variables of the reaction, but no factor of $i$ enters (\ref{e:UShadtens}) because Hermiticity implies the coefficients of the symmetric basis tensors are real.

Next, for the case where only $h_a$ is polarized, we now have $S_a^\mu$ along with $P_a^\mu$, $P_b^\mu$, $q^\mu$ as the vectors from which we can form our basis.  The antisymmetric tensors we can create that respect parity are given by 
\begin{align}
h_{a,1}^{A,\mu\nu},\,h_{a,2}^{A,\mu\nu},\, h_{a,3}^{A,\mu\nu} &= \epsilon^{qP_aP_bS_a}\,\{P_a^\mu P_b^\nu - P_a^\nu P_b^\mu,\,P_a^\mu q^\nu - P_a^\nu q^\mu,\, P_b^\mu q^\nu - P_b^\nu q^\mu\}\,, \label{e:haA1} \\
h_{a,4}^{A,\mu\nu},\,h_{a,5}^{A,\mu\nu},\, h_{a,6}^{A,\mu\nu} &= \{\epsilon^{\mu\nu S_a P_a},\, \epsilon^{\mu\nu S_aP_b},\, \epsilon^{\mu\nu S_a q}\}\,, \\
h_{a,7}^{A,\mu\nu},\,h_{a,8}^{A,\mu\nu} &= (\epsilon^{\mu q P_a P_b} q^\nu - \epsilon^{\nu q P_a P_b} q^\mu)\{q\cdot S_a, P_b \cdot S_a\}\,,\\
h_{a,9}^{A,\mu\nu},\,h_{a,10}^{A,\mu\nu} &= (\epsilon^{\mu q P_a P_b} P_a^\nu - \epsilon^{\nu q P_a P_b} P_a^\mu)\{q\cdot S_a, P_b \cdot S_a\}\,,\\
h_{a,11}^{A,\mu\nu},\,h_{a,12}^{A,\mu\nu} &= (\epsilon^{\mu q P_a P_b} P_b^\nu - \epsilon^{\nu q P_a P_b} P_b^\mu)\{q\cdot S_a, P_b \cdot S_a\}\,,\\
h_{a,13}^{A,\mu\nu} &= \epsilon^{\mu qP_aS_a}q^\nu -  \epsilon^{\nu qP_aS_a}q^\mu\,, \\
h_{a,14}^{A,\mu\nu} &= \epsilon^{\mu qS_aP_b}q^\nu -  \epsilon^{\nu qS_aP_b}q^\mu\,, \\
h_{a,15}^{A,\mu\nu} &= \epsilon^{\mu S_aP_aP_b}q^\nu -  \epsilon^{\nu qP_aP_b}q^\mu\,, \\
h_{a,16}^{A,\mu\nu} &= \epsilon^{\mu qP_aS_a}P_a^\nu -  \epsilon^{\nu qP_aS_a}P_a^\mu\,, \\
h_{a,17}^{A,\mu\nu} &= \epsilon^{\mu qS_aP_b}P_a^\nu -  \epsilon^{\nu qS_aP_b}P_a^\mu, \\
h_{a,18}^{A,\mu\nu} &= \epsilon^{\mu S_aP_aP_b}P_a^\nu -  \epsilon^{\nu S_aP_aP_b}P_a^\mu\,, \\
h_{a,19}^{A,\mu\nu} &= \epsilon^{\mu qP_aS_a}P_b^\nu -  \epsilon^{\nu qP_aS_a}P_b^\mu\,, \\
h_{a,20}^{A,\mu\nu} &= \epsilon^{\mu qS_aP_b}P_b^\nu -  \epsilon^{\nu qS_aP_b}P_b^\mu\,, \\
h_{a,21}^{A,\mu\nu} &= \epsilon^{\mu S_aP_aP_b}P_b^\nu -  \epsilon^{\nu S_aP_aP_b}P_b^\mu\,, \\
h_{a,22}^{A,\mu\nu} &= \epsilon^{\mu qP_aP_b}S_a^\nu -  \epsilon^{\nu qP_aP_b}S_a^\mu\,, \\
h_{a,23}^{A,\mu\nu},\,h_{a,24}^{A,\mu\nu},\, h_{a,25}^{A,\mu\nu} &= P_b\cdot S_a\,\{\epsilon^{\mu\nu P_a q},\,\epsilon^{\mu\nu P_b q},\,\epsilon^{\mu\nu P_a P_b}\}\,, \\
h_{a,26}^{A,\mu\nu},\,h_{a,27}^{A,\mu\nu},\, h_{a,28}^{A,\mu\nu} &= q\cdot S_a\,\{\epsilon^{\mu\nu P_a q},\,\epsilon^{\mu\nu P_b q},\,\epsilon^{\mu\nu P_a P_b}\}\,. \label{e:haA28}
\end{align}
Note that the subscript $a$ indicates that $h_a$ is polarized, and we have used the shorthand $\epsilon^{abcd} = \epsilon^{\mu\nu\rho\sigma}a_\mu b_\nu c_\rho d_\sigma$.  However, not all of the tensors (\ref{e:haA1})--(\ref{e:haA28}) are independent of each other.  We can use the identity
\begin{equation}
g^{\alpha\beta}\epsilon^{\mu\nu\rho\sigma} = g^{\mu\beta}\epsilon^{\alpha\nu\rho\sigma} + g^{\nu\beta}\epsilon^{\mu\alpha\rho\sigma} + g^{\rho\beta}\epsilon^{\mu\nu\alpha\sigma} + g^{\sigma\beta}\epsilon^{\mu\nu\rho\alpha}
\end{equation}
to show that 19 of them can be eliminated because they can be written in terms of the other 9 tensors.  For example,
\begin{align}
h_{a,1}^{A,\mu\nu} &= \epsilon^{\tau\gamma\rho\sigma}q_\tau P_{a,\gamma} P_{b,\rho} S_{a,\sigma}g^{\mu\alpha}P_{a,\alpha}P_b^\nu - (\mu\leftrightarrow \nu) \nonumber\\
&= q\cdot P_a\, \epsilon^{\mu P_aP_bS_a} + M_a^2\,\epsilon^{q\mu P_bS_a}P_b^\nu + P_a\cdot P_b\,\epsilon^{qP_a\mu S_a}P_b^\nu - (\mu\leftrightarrow \nu) \nonumber\\
&= q\cdot P_a\, h_{a,21}^{A,\mu\nu} + M_a^2\, h_{a,20}^{A,\mu\nu} + P_a\cdot P_b\, h_{a,19}^{A,\mu\nu}.
\end{align}
Through relations among the other tensors that can be established in a similar way, one finds that $h_{a,1}^{A,\mu\nu}, \,h_{a,2}^{A,\mu\nu}, \,h_{a,3}^{A,\mu\nu}, \,h_{a,7}^{A,\mu\nu},\,\dots,\,h_{a,22}^{A,\mu\nu}$ can be removed.  Now we impose {\it em} gauge invariance on the remaining tensors.  This leads to the following set:
\begin{align}
t_{a,1}^{\prime  A,\mu\nu},\,t_{a,2}^{\prime A,\mu\nu},\, t_{a,3}^{\prime A,\mu\nu} &= \{\tilde{\epsilon}^{\,\mu\nu S_aP_a} + \epsilon^{\mu\nu S_aP_a},\,\tilde{\epsilon}^{\,\mu\nu S_aP_b} + \epsilon^{\mu\nu S_aP_b},\,\epsilon^{\mu\nu S_aq} \}\,, \\
t_{a,4}^{\prime A,\mu\nu},\,t_{a,5}^{\prime  A,\mu\nu},\, t_{a,6}^{\prime  A,\mu\nu} &= P_b\cdot S_a\,\{ \epsilon^{\mu\nu P_aq},\,\epsilon^{\mu\nu P_bq},\, \tilde{\epsilon}^{\,\mu\nu P_aP_b} +\epsilon^{\mu\nu P_aP_b}\}\,,\\
t_{a,7}^{\prime A,\mu\nu},\,t_{a,8}^{\prime  A,\mu\nu},\, t_{a,9}^{\prime  A,\mu\nu} &= q\cdot S_a\,\{ \epsilon^{\mu\nu P_aq},\,\epsilon^{\mu\nu P_bq},\, \tilde{\epsilon}^{\,\mu\nu P_aP_b} +\epsilon^{\mu\nu P_aP_b}\}\,,
\end{align}
where
\begin{equation}
\tilde{\epsilon}^{\,\mu\nu ab} = \frac{q^\mu \epsilon^{\nu q ab} - q^\nu \epsilon^{\mu q ab}} {q^2}\,. 
\end{equation}
We have used a $t^{\prime}$ for these tensors because it turns out once we contract them with $\epsilon^{\mu\nu ll^{\prime}}$ (i.e., calculate the cross section), there are redundant contributions.  For example,
\begin{align}
\epsilon_{\mu\nu}^{\;\;\;\;ll^{\prime}}t_{a,1}^{\prime A,\mu\nu} = q\cdot P_a\,(l-l^{\prime})\cdot S_a + q\cdot S_a\,(l^{\prime}-l)\cdot P_a = -\frac{q\cdot P_a} {q^2}\,\epsilon_{\mu\nu}^{\;\;\;\;ll^{\prime}}t_{a,3}^{\prime A,\mu\nu} + \frac{1} {q^2}\,\epsilon_{\mu\nu}^{\;\;\;\;ll^{\prime}}t_{a,7}^{\prime  A,\mu\nu}.
\end{align}
In this regard the analysis with polarized leptons differs from one with unpolarized leptons, where no such repetition of terms occurs.  In the end, one can discard 4 tensors, and we choose to eliminate $t_{a,1}^{\prime  A,\mu\nu}, \,t_{a,2}^{\prime  A,\mu\nu},\,t_{a,6}^{\prime  A,\mu\nu}$, $t_{a,9}^{\prime\,  A,\mu\nu}$.  This leaves us with our final form for the single-polarized antisymmetric piece of the hadronic tensor:
\begin{equation}
W_a^{A,\mu\nu} = \sum_{i=1}^5 iV^A_{a,i} \, t_{a,i}^{A,\mu\nu}, \label{e:aAhadtens}
\end{equation}
where
\begin{align}
t_{a,1}^{A,\mu\nu} &= \epsilon^{\mu\nu S_aq}\,,\\
t_{a,2}^{A,\mu\nu},\, t_{a,3}^{A,\mu\nu} &= P_b\cdot S_a \, \{\epsilon^{\mu\nu P_aq}, \epsilon^{\mu\nu P_bq}\}\,,\\
t_{a,4}^{A,\mu\nu},\, t_{a,5}^{A,\mu\nu} &=  q\cdot S_a\,\{\epsilon^{\mu\nu P_aq},\,\epsilon^{\mu\nu P_bq}\}\,.
\end{align}
Likewise, for the case where only $h_b$ is polarized, we have
\begin{equation}
W_b^{A,\mu\nu} = \sum_{i=1}^5 iV^A_{b,i} \, t_{b,i}^{A,\mu\nu}, \label{e:bAhadtens}
\end{equation}
where
\begin{align}
t_{b,1}^{A,\mu\nu} &= \epsilon^{\mu\nu S_bq}\,,\\
t_{b,2}^{A,\mu\nu},\, t_{b,3}^{A,\mu\nu} &= P_a\cdot S_b \, \{\epsilon^{\mu\nu P_aq}, \epsilon^{\mu\nu P_bq}\}\,,\\
t_{b,4}^{A,\mu\nu},\, t_{b,5}^{A,\mu\nu} &=  q\cdot S_b\,\{\epsilon^{\mu\nu P_aq},\,\epsilon^{\mu\nu P_bq}\}\,.
\end{align}
As before, we just give the result from Ref.$\!$\cite{Arnold:2008kf} for the symmetric part of the hadronic tensor when only $h_a$ or $h_b$, respectively, is polarized: 
\begin{equation}
W_a^{S,\mu\nu} = \sum_{i=1}^8 V^S_{a,i} \, t_{a,i}^{S,\mu\nu}, \label{e:aShadtens}
\end{equation}
where
\begin{align}
t_{a,1}^{S,\mu\nu},\dots,\, t_{a,4}^{S,\mu\nu}&= \epsilon^{S_aqP_aP_b}\left\{g^{\mu\nu}-\frac{q^\mu q^\nu} {q^2},\, \tilde{P}_a^\mu \tilde{P}_a^\nu,\,\tilde{P}_b^\mu \tilde{P}_b^\nu,\,\tilde{P}_a^\mu\tilde{P}_b^\nu + \tilde{P}_a^\nu\tilde{P}_b^\mu\right\}\,,\\
t_{a,5}^{S,\mu\nu},\,t_{a,6}^{S,\mu\nu} &= \{S_a\cdot q,\,S_a\cdot P_b\}\,(\epsilon^{\mu qP_aP_b}\tilde{P}_a^\nu + \epsilon^{\nu qP_aP_b}\tilde{P}_a^\mu)\,,\\
t_{a,7}^{S,\mu\nu},\,t_{a,8}^{S,\mu\nu} &= \{S_a\cdot q,\,S_a\cdot P_b\}\,(\epsilon^{\mu qP_aP_b}\tilde{P}_b^\nu + \epsilon^{\nu qP_aP_b}\tilde{P}_b^\mu)\,,
\end{align}
and
\begin{equation}
W_b^{S,\mu\nu} = \sum_{i=1}^8 V^S_{b,i} \, t_{b,i}^{S,\mu\nu}, \label{e:bShadtens} 
\end{equation}
where
\begin{align}
t_{b,1}^{S,\mu\nu},\dots,\, t_{b,4}^{S,\mu\nu}&= \epsilon^{S_bqP_bP_a}\left\{g^{\mu\nu}-\frac{q^\mu q^\nu} {q^2},\, \tilde{P}_a^\mu \tilde{P}_a^\nu,\,\tilde{P}_b^\mu \tilde{P}_b^\nu,\,\tilde{P}_a^\mu\tilde{P}_b^\nu + \tilde{P}_a^\nu\tilde{P}_b^\mu\right\}\,,\\[-0.05cm]
t_{b,5}^{S,\mu\nu},\,t_{b,6}^{S,\mu\nu} &= \{S_b\cdot q,\,S_b\cdot P_a\}\,(\epsilon^{\mu qP_bP_a}\tilde{P}_a^\nu + \epsilon^{\nu qP_bP_a}\tilde{P}_a^\mu)\,,\\[-0.05cm]
t_{b,7}^{S,\mu\nu},\,t_{b,8}^{S,\mu\nu} &= \{S_b\cdot q,\,S_b\cdot P_a\}\,(\epsilon^{\mu qP_bP_a}\tilde{P}_b^\nu + \epsilon^{\nu qP_bP_a}\tilde{P}_b^\mu)\,.
\end{align}

Lastly, for the situation where both $h_a$ and $h_b$ are polarized, the vectors available to us are $S_a^\mu$, $S_b^\mu$, $P_a^\mu$, $P_b^\mu$, and $q^\mu$.  This leads to the following set of antisymmetric tensors consistent with the parity constraint:
\begin{align}
h_{ab,1}^{A,\mu\nu},\,h_{ab,2}^{A,\mu\nu},\, h_{ab,3}^{A,\mu\nu} &= S_a\cdot S_b\,\{q^\mu P_a^\nu-q^\nu P_a^\mu,\,q^\mu P_b^\nu - q^\nu P_b^\mu,\,P_a^\mu P_b^\nu - P_a^\nu P_b^\mu\}\,,\label{e:hab1}\\[0.05cm]
h_{ab,4}^{A,\mu\nu},\,h_{ab,5}^{A,\mu\nu},\, h_{ab,6}^{A,\mu\nu} &= S_a\cdot q \,S_b\cdot q\,\{q^\mu P_a^\nu-q^\nu P_a^\mu,\,q^\mu P_b^\nu - q^\nu P_b^\mu,\,P_a^\mu P_b^\nu - P_a^\nu P_b^\mu\}\,,\\
h_{ab,7}^{A,\mu\nu},\,h_{ab,8}^{A,\mu\nu},\, h_{ab,9}^{A,\mu\nu} &= S_a\cdot q\, S_b\cdot P_a\,\{q^\mu P_a^\nu-q^\nu P_a^\mu,\,q^\mu P_b^\nu - q^\nu P_b^\mu,\,P_a^\mu P_b^\nu - P_a^\nu P_b^\mu\}\,,\\
h_{ab,10}^{A,\mu\nu},\,h_{ab,11}^{A,\mu\nu},\, h_{ab,12}^{A,\mu\nu} &= S_b\cdot q\, S_a\cdot P_b\,\{q^\mu P_a^\nu-q^\nu P_a^\mu,\,q^\mu P_b^\nu - q^\nu P_b^\mu,\,P_a^\mu P_b^\nu - P_a^\nu P_b^\mu\}\,,\\
h_{ab,13}^{A,\mu\nu},\,h_{ab,14}^{A,\mu\nu},\, h_{ab,15}^{A,\mu\nu} &= S_a\cdot P_b \,S_b\cdot P_a\,\{q^\mu P_a^\nu-q^\nu P_a^\mu,\,q^\mu P_b^\nu - q^\nu P_b^\mu,\,P_a^\mu P_b^\nu - P_a^\nu P_b^\mu\}\,,\\
h_{ab,16}^{A,\mu\nu},\,h_{ab,17}^{A,\mu\nu},\, h_{ab,18}^{A,\mu\nu} &= S_a\cdot q\,\{S_b^\mu q^\nu - S_b^\nu q^\mu,\,S_b^\mu P_a^\nu - S_b^\nu P_a^\mu,\,S_b^\mu P_b^\nu - S_b^\nu P_b^\mu\}\,,\\[0.05cm]
h_{ab,19}^{A,\mu\nu},\,h_{ab,20}^{A,\mu\nu},\, h_{ab,21}^{A,\mu\nu} &= S_b\cdot q\,\{S_a^\mu q^\nu - S_a^\nu q^\mu,\,S_a^\mu P_a^\nu - S_a^\nu P_a^\mu,\,S_a^\mu P_b^\nu - S_a^\nu P_b^\mu\}\,,\\[0.05cm]
h_{ab,22}^{A,\mu\nu},\,h_{ab,23}^{A,\mu\nu},\, h_{ab,24}^{A,\mu\nu} &= S_a\cdot P_b\,\{S_b^\mu q^\nu - S_b^\nu q^\mu,\,S_b^\mu P_a^\nu - S_b^\nu P_a^\mu,\,S_b^\mu P_b^\nu - S_b^\nu P_b^\mu\}\,,\\[0.05cm]
h_{ab,25}^{A,\mu\nu},\,h_{ab,26}^{A,\mu\nu},\, h_{ab,27}^{A,\mu\nu} &= S_b\cdot P_a\,\{S_a^\mu q^\nu - S_a^\nu q^\mu,\,S_a^\mu P_a^\nu - S_a^\nu P_a^\mu,\,S_a^\mu P_b^\nu - S_a^\nu P_b^\mu\}\,,\\[0.05cm]
h_{ab,28}^{A,\mu\nu} &= S_a^\mu S_b^\nu - S_a^\nu S_b^\mu\,. \label{e:hab28}
\end{align}
Note that the subscript $ab$ now indicates both hadrons are polarized.  We are able to eliminate one tensor from Eqs.$\,$(\ref{e:hab1})--(\ref{e:hab28}) through use of the determinant identity
\begin{equation}
D^{\mu\alpha\beta\gamma\delta;\nu\bar{\alpha}\bar{\beta}\bar{\gamma}\bar{\delta}} = \left| \begin{array}{ccccc}
 g^{\mu\nu} & g^{\mu\bar{\alpha}} & g^{\mu\bar{\beta}} & g^{\mu\bar{\gamma}} & g^{\mu\bar{\delta}}\\
 g^{\alpha\nu} & g^{\alpha\bar{\alpha}} & g^{\alpha\bar{\beta}} & g^{\alpha\bar{\gamma}} & g^{\alpha\bar{\delta}}\\
 g^{\beta\nu} & g^{\beta\bar{\alpha}} & g^{\beta\bar{\beta}} & g^{\beta\bar{\gamma}} & g^{\beta\bar{\delta}} \\
 g^{\gamma\nu} & g^{\gamma\bar{\alpha}} & g^{\gamma\bar{\beta}} & g^{\gamma\bar{\gamma}} & g^{\gamma\bar{\delta}} \\
 g^{\delta\nu} & g^{\delta\bar{\alpha}} & g^{\delta\bar{\beta}} & g^{\delta\bar{\gamma}} & g^{\delta\bar{\delta}} \end{array} \right| = 0\,, \label{e:Det}
 \end{equation}
which was also used in Ref.$\!$\cite{Arnold:2008kf} for the symmetric case.  From (\ref{e:Det}) one easily obtains
 \begin{equation}
D_{\mu\alpha\beta\gamma\delta;\nu\bar{\alpha}\bar{\beta}\bar{\gamma}\bar{\delta}}\,(S_a^\alpha S_b^{\bar{\alpha}} - S_a^{\bar{\alpha}} S_b^{\alpha})\,q^\beta q^{\bar{\beta}} P_a^{\gamma} P_a^{\bar{\gamma}} P_b^\delta P_b^{\bar{\delta}} = 0\,,
 \end{equation}
which allows us to write $h_{ab,28}^{A,\mu\nu}$ as a linear combination of the other 27 tensors.  Here we refrain from explicitly giving the lengthy expression, but the main point is that we can remove $h_{ab,28}^{A,\mu\nu}$.  The identity (\ref{e:Det}) does not lead to any further relations among the tensors.  We now impose {\it em} gauge invariance on the tensors that remain in order to determine the final form of the double-polarized antisymmetric part of the hadronic tensor:
 \begin{equation}
W_{ab}^{A,\mu\nu} = \sum_{i=1}^{13} iV^A_{ab,i} \, t_{ab,i}^{A,\mu\nu}, \label{e:abAhadtens}
\end{equation}
where
\begin{align}
t_{ab,1}^{A,\mu\nu},\dots,\, t_{ab,5}^{A,\mu\nu} &= (\tilde{P}_a^\mu \tilde{P}_b^\nu - \tilde{P}_a^\nu \tilde{P}_b^\mu)\,\{S_a\cdot S_b, S_a\cdot q\,S_b\cdot q, \nonumber \\[-0.15cm]
	& \hspace{3.65cm} \, S_a\cdot q\,S_b\cdot P_a,\,S_b\cdot q\,S_a\cdot P_b,\,S_a\cdot P_b\, S_b\cdot P_a\}\,,\\
t_{ab,6}^{A,\mu\nu},\,t_{ab,7}^{A,\mu\nu} &= S_a\cdot q\,\{\tilde{S}_b^\mu \tilde{P}_a^\nu - \tilde{S}_b^\nu \tilde{P}_a^\mu,\, \tilde{S}_b^\mu \tilde{P}_b^\nu - \tilde{S}_b^\nu \tilde{P}_b^\mu\}\,, \\
t_{ab,8}^{A,\mu\nu},\,t_{ab,9}^{A,\mu\nu} &= S_b\cdot q\,\{\tilde{S}_a^\mu \tilde{P}_a^\nu - \tilde{S}_a^\nu \tilde{P}_a^\mu,\, \tilde{S}_a^\mu \tilde{P}_b^\nu - \tilde{S}_a^\nu \tilde{P}_b^\mu\}\,, \\
t_{ab,10}^{A,\mu\nu},\,t_{ab,11}^{A,\mu\nu} &= S_a\cdot P_b\,\{\tilde{S}_b^\mu \tilde{P}_a^\nu - \tilde{S}_b^\nu \tilde{P}_a^\mu,\, \tilde{S}_b^\mu \tilde{P}_b^\nu - \tilde{S}_b^\nu \tilde{P}_b^\mu\}\,, \\
t_{ab,12}^{A,\mu\nu},\,t_{ab,13}^{A,\mu\nu} &= S_b\cdot P_a\,\{\tilde{S}_a^\mu \tilde{P}_a^\nu - \tilde{S}_a^\nu \tilde{P}_a^\mu,\, \tilde{S}_a^\mu \tilde{P}_b^\nu - \tilde{S}_a^\nu \tilde{P}_b^\mu\}\,.
\end{align}
Again, we simply state the result for the symmetric part of the double-polarized hadronic tensor \cite{Arnold:2008kf}:
\begin{equation}
W_{ab}^{S,\mu\nu} = \sum_{i=1}^{28} V^S_{ab,i} \, t_{ab,i}^{S,\mu\nu}, \label{e:abShadtens}
\end{equation}
where
\begin{align}
t_{ab,1}^{S,\mu\nu},\dots,\, t_{ab,4}^{S,\mu\nu} &= S_a\cdot S_b\, \left\{g^{\mu\nu}-\frac{q^\mu q^\nu} {q^2},\, \tilde{P}_a^\mu \tilde{P}_a^\nu,\, \tilde{P}_b^\mu \tilde{P}_b^\nu,\, \tilde{P}_a^\mu \tilde{P}_b^\nu + \tilde{P}_a^\nu \tilde{P}_b^\mu \right\}, \\[0.1cm]
t_{ab,5}^{S,\mu\nu},\dots,\, t_{ab,8}^{S,\mu\nu} &= S_a\cdot q\, S_b\cdot q\, \left\{g^{\mu\nu}-\frac{q^\mu q^\nu} {q^2},\, \tilde{P}_a^\mu \tilde{P}_a^\nu,\, \tilde{P}_b^\mu \tilde{P}_b^\nu,\, \tilde{P}_a^\mu \tilde{P}_b^\nu + \tilde{P}_a^\nu \tilde{P}_b^\mu \right\}, \\[0.1cm]
t_{ab,9}^{S,\mu\nu},\dots,\, t_{ab,12}^{S,\mu\nu} &= S_a\cdot q\, S_b\cdot P_a\, \left\{g^{\mu\nu}-\frac{q^\mu q^\nu} {q^2},\, \tilde{P}_a^\mu \tilde{P}_a^\nu,\, \tilde{P}_b^\mu \tilde{P}_b^\nu,\, \tilde{P}_a^\mu \tilde{P}_b^\nu + \tilde{P}_a^\nu \tilde{P}_b^\mu \right\}, \\[0.1cm]
t_{ab,13}^{S,\mu\nu},\dots,\, t_{ab,16}^{S,\mu\nu} &= S_b\cdot q\, S_a\cdot P_b\, \left\{g^{\mu\nu}-\frac{q^\mu q^\nu} {q^2},\, \tilde{P}_a^\mu \tilde{P}_a^\nu,\, \tilde{P}_b^\mu \tilde{P}_b^\nu,\, \tilde{P}_a^\mu \tilde{P}_b^\nu + \tilde{P}_a^\nu \tilde{P}_b^\mu \right\}, \\[0.1cm]
t_{ab,17}^{S,\mu\nu},\dots,\, t_{ab,20}^{S,\mu\nu} &= S_a\cdot P_b\, S_b\cdot P_a\left\{g^{\mu\nu}-\frac{q^\mu q^\nu} {q^2},\, \tilde{P}_a^\mu \tilde{P}_a^\nu,\, \tilde{P}_b^\mu \tilde{P}_b^\nu,\, \tilde{P}_a^\mu \tilde{P}_b^\nu + \tilde{P}_a^\nu \tilde{P}_b^\mu \right\}, \\[0.1cm]
t_{ab,21}^{S,\mu\nu},\, t_{ab,22}^{S,\mu\nu} &= S_a\cdot q\,\{\tilde{S}_b^\mu \tilde{P}_a^\nu + \tilde{S}_b^\nu \tilde{P}_a^\mu,\, \tilde{S}_b^\mu\tilde{P}_b^\nu + \tilde{S}_b^\nu\tilde{P}_b^\mu\}\,, \\
t_{ab,23}^{S,\mu\nu},\, t_{ab,24}^{S,\mu\nu} &= S_b\cdot q\,\{\tilde{S}_a^\mu \tilde{P}_b^\nu + \tilde{S}_a^\nu \tilde{P}_b^\mu,\, \tilde{S}_a^\mu\tilde{P}_a^\nu + \tilde{S}_a^\nu\tilde{P}_a^\mu\}\,, \\
t_{ab,25}^{S,\mu\nu},\, t_{ab,26}^{S,\mu\nu} &= S_a\cdot P_b\,\{\tilde{S}_b^\mu \tilde{P}_a^\nu + \tilde{S}_b^\nu \tilde{P}_a^\mu,\, \tilde{S}_b^\mu\tilde{P}_b^\nu + \tilde{S}_b^\nu\tilde{P}_b^\mu\}\,, \\
t_{ab,27}^{S,\mu\nu},\, t_{ab,28}^{S,\mu\nu} &= S_b\cdot P_a\,\{\tilde{S}_a^\mu \tilde{P}_b^\nu + \tilde{S}_a^\nu \tilde{P}_b^\mu,\, \tilde{S}_a^\mu\tilde{P}_a^\nu + \tilde{S}_a^\nu\tilde{P}_a^\mu\}\,.
\end{align}

Thus, the total hadronic tensor for the pure electromagnetic case reads
\begin{equation}
W^{\mu\nu}_{\gamma\gamma} = W_{U}^{A,\mu\nu} + W_{U}^{S,\mu\nu} + W_{a}^{A,\mu\nu} + W_{b}^{A,\mu\nu} + W_{a}^{S,\mu\nu} + W_{b}^{S,\mu\nu} + W_{ab}^{A,\mu\nu} + W_{ab}^{S,\mu\nu},
\label{e:hadtensFinal}
\end{equation}
where the individual terms are given by Eqs.$\,$(\ref{e:UAhadtens}),$\,$(\ref{e:UShadtens}),$\,$(\ref{e:aAhadtens}),$\,$(\ref{e:bAhadtens}),$\,$(\ref{e:aShadtens}),$\,$(\ref{e:bShadtens}), (\ref{e:abAhadtens}),$\,$(\ref{e:abShadtens}), respectively, and 72 structure functions enter into the result (24 associated with antisymmetric basis tensors and 48 with symmetric).  If we only consider $W_{U}^{A} + W_{U}^{S} + W_{a}^{A} + W_{a}^{S}$, then 18 structure functions show up, which is exactly the same number obtained in, e.g., SIDIS when one allows for beam and target polarization \cite{Diehl:2005pc, Bacchetta:2006tn}.  We mention again that, using a diagrammatic approach, the hadronic tensor was written down before in Ref.$\!$\cite{Boer:1997mf} but only within a specific frame and just up to terms of twist-3 accuracy.  Our analysis has provided a decomposition of the hadronic tensor valid to any twist and of use in any frame.
%
%
%
\section{Reference frames and the cross section}
\label{s:CSnRef} 

\noindent We are now in a position to calculate the general form of the cross section by contracting the leptonic tensor (\ref{e:leptens2}) with the hadronic tensor (\ref{e:hadtensFinal}) (cf.~Eq.$\,$(\ref{e:epluseminusCross})).  However, in order for one to obtain a general angular distribution of this cross section, a reference frame must be chosen.  Here we have in mind that the two hadrons are almost back-to-back.  In the following we specify a di-lepton rest frame that is similar to the Collins-Soper (CS) frame \cite{Collins:1977iv} and define the hadronic {\it cm} frame.  The former was introduced (along with another di-lepton rest frame akin to the Gottfried-Jackson (GJ) frame \cite{Gottfried:1964nx}) in the context of $e^+e^- \!\rightarrow h_a h_b X$ in Refs.$\!$\cite{Boer:1997mf, Boer:2008fr}.  Both frames are also widely used when studying the Drell-Yan process.  (For $e^+e^-$ collisions, it might seem more natural to call the ``di-lepton rest frame'' a ``leptonic {\it cm} frame'' and the ``hadronic {\it cm} frame'' a ``di-hadron rest frame.''  However, in order to avoid confusion, we will stick with the terminology from Drell-Yan, especially since we refer to the CS (and GJ) frames.)  We mention that the experimental analyses of $e^+e^-\! \rightarrow h_a h_b X$ performed by Belle and BABAR \cite{Abe:2005zx, TheBABAR:2013yha} are done twice:~once in the GJ frame, and once in the so-called ``thrust-axis'' frame \cite{Boer:2008fr, Anselmino:2007fs, Bonivento:1995}.  The latter cannot be related to the CS frame and actually requires the detection of an additional jet in the final state.  Nevertheless, the cross section takes on its most compact and transparent form when written in terms of angles from a di-lepton rest frame.  In this case we will use angles from the CS frame but will explain in Sect.$\,$\ref{s:SFT2} how one can easily write down the cross section (and find values for the structure functions) at twist-2 in terms of angles from the GJ frame, and, thus, make an exact connection to experiment.
\begin{figure}[t]
\begin{center}
\includegraphics[width=11cm]{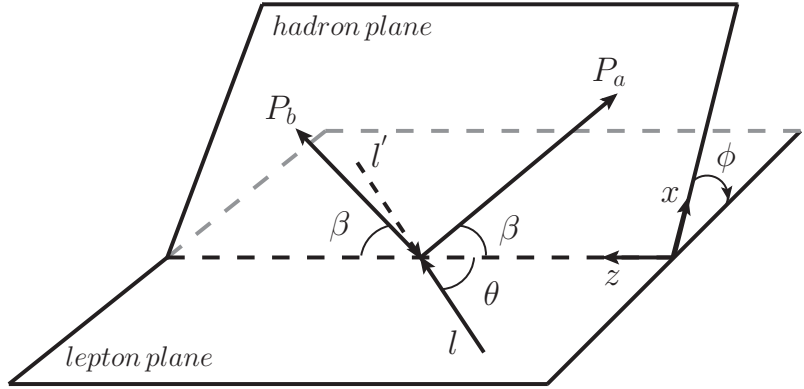}
\caption[]{Analogue of the Collins-Soper frame for $e^+e^-\!\!\rightarrow h_a h_b\, X$.  The incoming electron makes an angle $\theta$ w.r.t.~the $+z$-axis, and the plane spanned by the outgoing hadrons forms an angle $\phi$ w.r.t.~the lepton plane.  Note that both hadrons form the same angle $\beta$ w.r.t~the $+z$-axis.}
 \label{f:CSframe}
\end{center}
\end{figure}

\begin{figure}[t]
\begin{center}
\includegraphics[width=9cm]{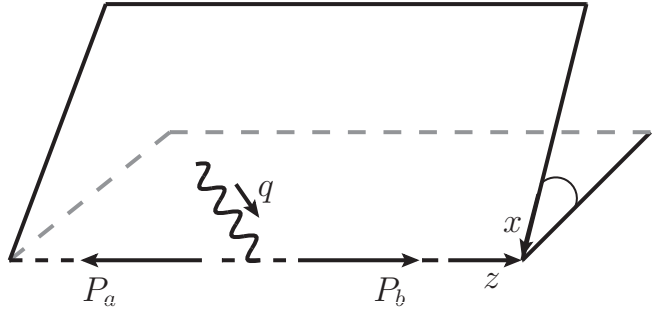}
\caption[]{Hadronic center-of-mass frame for $e^+e^-\!\!\rightarrow h_a h_b\, X$.  The hadron $h_b$ ($h_a$) moves along the $+z$-axis ($-z$-axis), and the transverse momentum of the virtual photon defines the $+x$-axis.}
 \label{f:hadroncm_frame}
\end{center}
\end{figure}

The analogue of the CS frame is shown in Fig.$\,$\ref{f:CSframe}, and the components of $l^\mu$, $l^{\prime \mu}$, $P_a^\mu$, $P_b^\mu$, $q^\mu$ in this frame are given by
\begin{align}
l_{CS}^\mu &= \frac{q} {2}\, (1,\,\sin\theta\cos\phi,\,\sin\theta\sin\phi,\,\cos\theta)\,, \label{e:lCS} \\
l_{CS}^{\prime\mu} &= \frac{q} {2}\,(1,\,-\sin\theta\cos\phi,\,-\sin\theta\sin\phi,\,-\cos\theta)\,, \label{e:lprimeCS}\\
P_{a,CS}^\mu &\approx P_{a,CS}^0\,(1,\,\sin\beta,\,0,\,-\cos\beta)\,,\label{e:PaCS1}\\
P_{b,CS}^\mu &\approx P_{b,CS}^0\,(1,\,\sin\beta,\,0,\,\cos\beta)\,,\label{e:PbCS1}\\
q_{CS}^\mu &= (q,\,0,\,0,\,0)\,,
\end{align}
where for $P_a$ and $P_b$ we have neglected the masses of the hadrons.  On the other hand, the components of $P_a^\mu$, $P_b^\mu$, $q^\mu$ in the hadronic {\it cm} frame shown in Fig.$\,$\ref{f:hadroncm_frame} are given by\vspace{-0.2cm}
\begin{align}
P_{a,cm}^\mu &\approx P_{a,cm}^0\,(1,\,0,\,0,\,-1)\,,\label{e:Pacm}\\
P_{b,cm}^\mu &\approx P_{b,cm}^0\,(1,\,0,\,0,\,1)\,,\\
q_{cm}^\mu &= (q^0_{cm},\,q_{\perp,cm},\,0,\,q_{L,cm})\,. \label{e:qcm}
\end{align}
Note that we can fix the transverse momentum of the virtual photon to be along the $+x$-axis without loss of generality.  

One can obtain the Lorentz transformation $M^\mu_{\;\;\,\nu}$ from the hadronic {\it cm} frame to the CS frame through a boost along the $+z$-axis that eliminates $q_{L,cm}$ followed by a boost along the $+x$-axis that removes $q_{\perp,cm}$.  The result is\vspace{0.2cm}
\begin{equation}
M^\mu_{\;\;\,\nu} = \frac{1} {q}\left(\!\begin{array} {cccc}
q_{cm}^0 & -q_{\perp,cm} & 0 & -q_{L,cm}\\[0.1cm]
-q_{cm}^0 \sin\xi & q/\cos\xi & 0 & q_{L,cm}\sin\xi \\[0.1cm]
0 & 0 & q & 0\\[0.1cm]
-q_{L,cm}\cos\xi & 0 & 0 & q_{cm}^0\cos\xi
\end{array}\!\right),
\label{e:cmCStrans} \vspace{0.2cm}
\end{equation}
where $\cos\xi = 1/\sqrt{1+\rho^2}$ and $\sin\xi = \rho/\sqrt{1+\rho^2}$ with $\rho = q_{\perp,cm}/q$.  (Note that a boost first along the $+x$-axis and then along the $+z$-axis does {\it not} take us to the CS frame.)  Applying the matrix (\ref{e:cmCStrans}) to Eqs.$\,$(\ref{e:Pacm})--(\ref{e:qcm}) gives us
\begin{align}
P_{a,CS}^\mu &\approx \frac{P_{a,cm}^0} {q}(q_{cm}^0+q_{L,cm})\,(1,\,-\sin\xi,\,0,\,-\cos\xi)\,,\label{e:PaCS2}\\
P_{b,CS}^\mu &\approx \frac{P_{b,cm}^0} {q}(q_{cm}^0-q_{L,cm})\,(1,\,-\sin\xi,\,0,\,\cos\xi)\,,\label{e:PbCS2}\\
q_{CS}^\mu &= (q,\,0,\,0,\,0)\,.
\end{align}
Comparing Eqs.$\,$(\ref{e:PaCS2}),$\,$(\ref{e:PbCS2}) to Eqs.$\,$(\ref{e:PaCS1}),$\,$(\ref{e:PbCS1}) allows us to make the identifications
\begin{align}
P_{a,CS}^0 &= \frac{P_{a,cm}^0} {q}(q_{cm}^0+q_{L,cm})\,,\\[0.1cm]
P_{b,CS}^0 &= \frac{P_{b,cm}^0} {q}(q_{cm}^0-q_{L,cm})\,,\\
\beta &= -\xi\,. 
\end{align}
Likewise, the inverse Lorentz transformation that takes us from the CS frame to the hadronic {\it cm} frame reads\vspace{0.2cm}
\begin{equation}
(M^{-1})^\mu_{\;\;\nu} = \frac{1} {q}\left(\!\begin{array} {cccc}
q_{cm}^0 & q_{cm}^0 \sin\xi & 0 & q_{L,cm}\cos\xi\\[0.1cm]
q_{\perp,cm} & q/\cos\xi & 0 & 0 \\[0.1cm]
0 & 0 & q & 0\\[0.1cm]
q_{L,cm} & q_{L,cm}\sin\xi & 0 & q_{cm}^0\cos\xi
\end{array}\!\right).
\label{e:CScmtrans} \vspace{0.2cm}
\end{equation}
Applying the matrix (\ref{e:CScmtrans}) to Eqs.$\,$(\ref{e:lCS}),$\,$(\ref{e:lprimeCS}) gives us\vspace{0.2cm}
\begin{align}
l_{cm}^\mu &= \frac{1} {2}\left(\!\begin{array} {c}
q_{cm}^0(1+\sin\xi\sin\theta\cos\phi) + q_{L,cm}\cos\xi\cos\theta\\
q_{\perp,cm} +q\sin\theta\cos\phi/\cos\xi \\
q\sin\theta\sin\phi\\
q_{L,cm}(1+\sin\xi\sin\theta\cos\phi) + q_{cm}^0\cos\xi\cos\theta
\end{array}\!\right),\label{e:lcm}\\[0.5cm]
l_{cm}^{\prime \mu} &= \frac{1} {2}\left(\!\begin{array} {c}
q_{cm}^0(1-\sin\xi\sin\theta\cos\phi) - q_{L,cm}\cos\xi\cos\theta\\
q_{\perp,cm} -q\sin\theta\cos\phi/\cos\xi \\
-q\sin\theta\sin\phi\\
q_{L,cm}(1-\sin\xi\sin\theta\cos\phi) - q_{cm}^0\cos\xi\cos\theta
\end{array}\!\right)\label{e:lprimecm}.
\end{align} 
One can also write the covariant spin vectors $S_a^\mu$, $S_b^\mu$ in the hadronic {\it cm} frame:\vspace{0.2cm}
\begin{align}
S_{a,cm}^\mu &= \left(\Lambda_{a,cm}\,\frac{|\vec{P}_{a,cm}|} {M_a},\,|\vec{S}_{a\perp,cm}|\cos\phi_a,\,|\vec{S}_{a\perp,cm}|\sin\phi_a,\,-\Lambda_{a,cm}\frac{P_{a,cm}^0} {M_a}\right),\label{e:haspincm}\\[0.3cm]
S_{b,cm}^\mu &= \left(\Lambda_{b,cm}\,\frac{|\vec{P}_{b,cm}|} {M_b},\,|\vec{S}_{b\perp,cm}|\cos\phi_b,\,|\vec{S}_{b\perp,cm}|\sin\phi_b,\,\Lambda_{b,cm}\frac{P_{b,cm}^0} {M_b}\right), \label{e:hbspincm}
\end{align}
where $\Lambda_{a(b),cm}$ and $\vec{S}_{a(b)\perp,cm}$ are the helicity and transverse spin, respectively, for $h_a$ ($h_b$), and the azimuthal angle of $\vec{S}_{a(b)\perp,cm}$ w.r.t.~$\vec{q}_{\perp,cm}$ is given by $\phi_{a(b)}$. 

We are finally in a position to give the general angular decomposition of the first term in the cross section (\ref{e:epluseminusCross}).  The computation involves the contraction of two Lorentz tensors (i.e., $L^{\mu\nu}_{\gamma\gamma},\,W^{\mu\nu}_{\gamma\gamma}$), which we can perform in any frame.  We choose the hadronic {\it cm} frame in part because the FFs are understood to be defined in a frame where the outgoing hadron has no transverse momentum.  Therefore, it will be necessary to use this frame when we calculate the structure functions in terms of transverse momentum dependent (TMD) FFs up to twist-2 accuracy in Sect.$\,$\ref{s:SFT2}.  Moreover, the result retains a compact and transparent form when expressed through the CS angles $\theta$, $\phi$.  Therefore, we will use the expressions (\ref{e:lcm}),$\,$(\ref{e:lprimecm}) for the lepton momenta.  In the end, one finds\vspace{0.2cm}
\begin{align}
4&\frac{P_a^0P_b^0d\sigma_{em}} {d^3\vec{P}_ad^3\vec{P}_b} = \frac{\alpha_{em}^2} {q^4}\, \times\nonumber \\[0.2cm]
&\;\Bigg\{\bigg\{\!\left[(1+\cos^2\theta)F_{UU}^1 + (1-\cos^2\theta)F_{UU}^3 + (\sin2\theta\cos\phi)F_{UU}^{\cos\phi} + (\sin^2\theta\cos2\phi)F_{UU}^{\cos2\phi}\right]\nonumber\\[0.05cm]
&\;\;\;\;\;\; +\,\Lambda_{a}\left[(\sin^2\theta\sin2\phi)F_{LU}^{\sin2\phi}+(\sin2\theta\sin\phi)F_{LU}^{\sin\phi}\right]\nonumber\\[0.05cm]
&\;\;\;\;\;\; +\,|\vec{S}_{a\perp}|\left[\sin\phi_a\left((1+\cos^2\theta)F_{TU}^1 + (1-\cos^2\theta)F_{TU}^3 +(\sin2\theta\cos\phi)F_{TU}^{\cos\phi} \right.\right.\nonumber\\[-0.3cm]
	&\hspace{10.77cm}\left. +\, (\sin^2\theta\cos2\phi)F_{TU}^{\cos2\phi}\right)\nonumber\\[0.05cm]
&\hspace{2.1cm}\left.+\,\cos\phi_a\left((\sin^2\theta\sin2\phi)F_{TU}^{\sin2\phi}+(\sin2\theta\sin\phi)F_{TU}^{\sin\phi}\right)\right]\nonumber\\[0.05cm]
&\;\;\;\;\;\; +\,\Lambda_{b}\left[(\sin^2\theta\sin2\phi)F_{UL}^{\sin2\phi}+(\sin2\theta\sin\phi)F_{UL}^{\sin\phi}\right]\nonumber\\[0.05cm]
&\;\;\;\;\;\; +\,|\vec{S}_{b\perp}|\left[\sin\phi_b\left((1+\cos^2\theta)F_{UT}^1 + (1-\cos^2\theta)F_{UT}^3 +(\sin2\theta\cos\phi)F_{UT}^{\cos\phi} \right.\right.\nonumber\\[-0.25cm]
	&\hspace{10.77cm}\left. +\, (\sin^2\theta\cos2\phi)F_{UT}^{\cos2\phi}\right)\nonumber\\[0.05cm]
&\hspace{2.1cm}\left.+\,\cos\phi_b\left((\sin^2\theta\sin2\phi)F_{UT}^{\sin2\phi}+(\sin2\theta\sin\phi)F_{UT}^{\sin\phi}\right)\right]\nonumber\\[0.05cm]
&\;\;\;\;\;\; +\,\Lambda_{a}\Lambda_{b}\left[(1+\cos^2\theta)F_{LL}^1 + (1-\cos^2\theta)F_{LL}^3 + (\sin2\theta\cos\phi)F_{LL}^{\cos\phi} \right.\nonumber\\[-0.2cm]
	&\hspace{10.82cm}\left.+\, (\sin^2\theta\cos2\phi)F_{LL}^{\cos2\phi}\right]\nonumber\\[0.05cm]
&\;\;\;\;\;\; +\,\Lambda_{a}|\vec{S}_{b\perp}|\left[\cos\phi_b\left((1+\cos^2\theta)F_{LT}^1 + (1-\cos^2\theta)F_{LT}^3 +(\sin2\theta\cos\phi)F_{LT}^{\cos\phi} \right.\right.\nonumber\\[-0.2cm]
	&\hspace{10.77cm}\left. +\, (\sin^2\theta\cos2\phi)F_{LT}^{\cos2\phi}\right)\nonumber\\[0.05cm]
&\hspace{2.7cm}\left.+\,\sin\phi_b\left((\sin^2\theta\sin2\phi)F_{LT}^{\sin2\phi}+(\sin2\theta\sin\phi)F_{LT}^{\sin\phi}\right)\right]\nonumber\\[0.05cm]
&\;\;\;\;\;\; +\,|\vec{S}_{a\perp}|\Lambda_{b}\left[\cos\phi_a\left((1+\cos^2\theta)F_{TL}^1 + (1-\cos^2\theta)F_{TL}^3 +(\sin2\theta\cos\phi)F_{TL}^{\cos\phi} \right.\right.\nonumber\\[-0.2cm]
	&\hspace{10.77cm}\left. +\, (\sin^2\theta\cos2\phi)F_{TL}^{\cos2\phi}\right)\nonumber\\[0.05cm]
&\hspace{2.7cm}\left.+\,\sin\phi_a\left((\sin^2\theta\sin2\phi)F_{TL}^{\sin2\phi}+(\sin2\theta\sin\phi)F_{TL}^{\sin\phi}\right)\right]\nonumber\\[0.05cm]
&\;\;\;\;\;\; +\,|\vec{S}_{a\perp}||\vec{S}_{b\perp}|\left[\cos(\phi_a+\phi_b)\Big((1+\cos^2\theta)F_{TT}^1 + (1-\cos^2\theta)F_{TT}^3 \right.\nonumber\\[-0.2cm]
	&\hspace{7cm}\left. +\,  (\sin2\theta\cos\phi)F_{TT}^{\cos\phi} + (\sin^2\theta\cos2\phi)F_{TT}^{\cos2\phi}\right)\nonumber\\[0.05cm]
&\hspace{2.95cm}+\,\cos(\phi_a-\phi_b)\Big((1+\cos^2\theta)\bar{F}_{TT}^1 + (1-\cos^2\theta)\bar{F}_{TT}^3 \nonumber\\[-0.2cm]
	&\hspace{7cm}\left. +\,  (\sin2\theta\cos\phi)\bar{F}_{TT}^{\cos\phi} + (\sin^2\theta\cos2\phi)\bar{F}_{TT}^{\cos2\phi}\right)\nonumber\\[0.05cm]
&\hspace{2.95cm}+\,\sin(\phi_a+\phi_b)\left((\sin^2\theta\sin2\phi)F_{TT}^{\sin2\phi} + (\sin2\theta\sin\phi)F_{TT}^{\sin\phi} \right)\nonumber\\[0.05cm]
&\hspace{2.95cm}\left.+\,\sin(\phi_a-\phi_b)\left((\sin^2\theta\sin2\phi)\bar{F}_{TT}^{\sin2\phi} + (\sin2\theta\sin\phi)\bar{F}_{TT}^{\sin\phi} \right)\right]\!\bigg\}\nonumber\\[0.05cm]
&\;-2\lambda_e\,\bigg\{\!\left[(\sin\theta\sin\phi)G_{UU}^{\sin\phi}\right]\nonumber\\[0.05cm]
&\hspace{1.5cm}+\,\Lambda_{a}\left[(\cos\theta)G_{LU}^2 + (\sin\theta\cos\phi)G_{LU}^{\cos\phi}\right]\nonumber\\[0.05cm]
&\hspace{1.5cm}+\,|\vec{S}_{a\perp}|\left[\cos\phi_a\left((\cos\theta)\bar{G}_{TU}^2 + (\sin\theta\cos\phi)G_{TU}^{\cos\phi}\right)+\, \sin\phi_a\left((\sin\theta\sin\phi)G_{TU}^{\sin\phi}\right)\right]\nonumber\\[0.05cm]
&\hspace{1.5cm}+\,\Lambda_{b}\left[(\cos\theta)G_{UL}^2 + (\sin\theta\cos\phi)G_{UL}^{\cos\phi}\right]\nonumber\\[0.05cm]
&\hspace{1.5cm}+\,|\vec{S}_{b\perp}|\left[\cos\phi_b\left((\cos\theta)\bar{G}_{UT}^2 + (\sin\theta\cos\phi)G_{UT}^{\cos\phi}\right) +\, \sin\phi_b\left((\sin\theta\sin\phi)G_{UT}^{\sin\phi}\right)\right]\nonumber\\[0.05cm]
&\hspace{1.5cm}+\, \Lambda_{a}\Lambda_{b}\left[(\sin\theta\sin\phi)G_{LL}^{\sin\phi}\right]\nonumber\\[0.05cm]
&\hspace{1.5cm}+\, \Lambda_{a}|\vec{S}_{b\perp}|\left[\cos\phi_b\left((\sin\theta\sin\phi)G_{LT}^{\sin\phi}\right)+\sin\phi_b\left((\cos\theta)\bar{G}_{LT}^2+(\sin\theta\cos\phi)G_{LT}^{\cos\phi}\right)\right]\nonumber\\[0.05cm]
&\hspace{1.5cm}+\, |\vec{S}_{a\perp}|\Lambda_{b}\left[\cos\phi_a\left((\sin\theta\sin\phi)G_{TL}^{\sin\phi}\right)+ \sin\phi_a\left((\cos\theta)\bar{G}_{TL}^2+(\sin\theta\cos\phi)G_{TL}^{\cos\phi}\right)\right]\nonumber\\[0.05cm]
&\hspace{1.5cm}+\, |\vec{S}_{a\perp}||\vec{S}_{b\perp}|\left[\cos(\phi_a+\phi_b)\left((\sin\theta\sin\phi)G_{TT}^{\sin\phi}\right)\right.\nonumber\\[0.05cm]
&\hspace{3.8cm}+\, \cos(\phi_a-\phi_b)\left((\sin\theta\sin\phi)\bar{G}_{TT}^{\sin\phi}\right)\nonumber\\[0.05cm]
&\hspace{3.8cm}+\,\sin(\phi_a+\phi_b)\left((\cos\theta)\ddot{G}_{TT}^2+(\sin\theta\cos\phi)G_{TT}^{\cos\phi}\right)\nonumber\\[0.05cm]
&\hspace{3.8cm}\left.+\, \sin(\phi_a-\phi_b)\left((\cos\theta)\hat{G}_{TT}^2+(\sin\theta\cos\phi)\bar{G}_{TT}^{\cos\phi}\right)\right]\!\bigg\}\Bigg\}\,. \label{e:angdecomp}
\end{align}
We mention that the notation used for some of the structure functions might seem ``weird'' in that, e.g., $F_{UU}^3$ shows up but there is no $F_{UU}^2$, or $\bar{G}_{TU}^2$ looks like an odd naming choice.  However, this notation provides a consistency among results when we discuss the electroweak case in Sect.$\,$\ref{s:SFT2}.  Note that 72 structure functions appear in Eq.$\,$(\ref{e:angdecomp}), which is exactly the same number that we wrote down when we decomposed the hadronic tensor in Sect.$\,$\ref{s:DecompHad}.  Again these are real valued functions that depend on scalar variables of the reaction.  The terms involving unpolarized leptons have the same angular structure as the Drell-Yan case analyzed in \cite{Arnold:2008kf} (see also \cite{Ralston:1979ys, Tangerman:1994eh, Pire:1983tv, Boer:1997nt, Boglione:2011zw}).  Also, as Ref.$\!$\cite{Arnold:2008kf} emphasized, the angular distribution of the cross section is the same for any di-lepton rest frame.  That is, the angles $\phi$ and $\theta$ are the azimuthal angle of the hadron plane and polar angle of the incoming electron, respectively, of whichever di-lepton rest frame one chooses, not just the CS frame.  Furthermore, the spin components can be understood in different frames, not just the hadronic {\it cm} frame.  Of course, the structure functions will take on different values in each frame.
%
%
%
\section{Structure functions at twist-2}
\label{s:SFT2} 

\begin{figure}[t]
\begin{center}
\includegraphics[width=15.35cm]{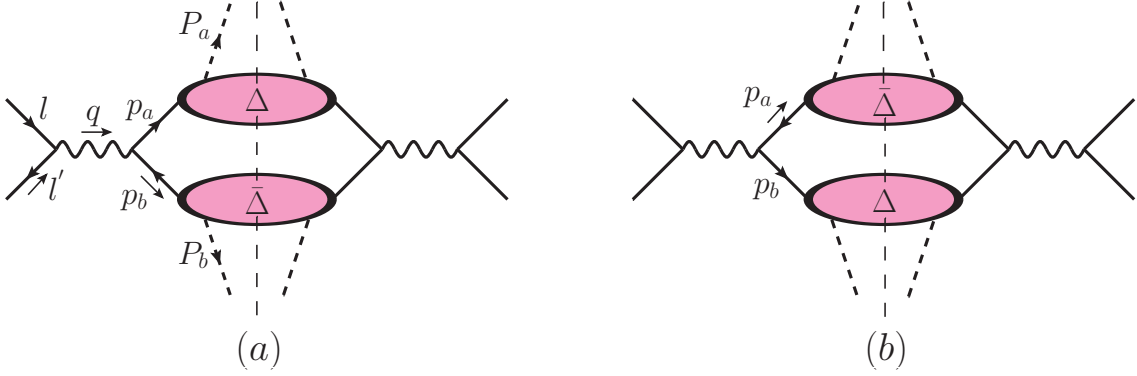}
\caption[]{Cross section for $e^+e^-\!\!\rightarrow h_ah_b\, X$ in a partonic description for $q_{\perp,cm} \ll q$.  The virtual photon decays into a quark-antiquark pair with (a) the quark (antiquark) fragmenting into $h_a$ ($h_b$) or (b) the antiquark (quark) fragmenting into $h_a$ ($h_b$).}
 \label{f:epluseminusPartonFact}
\end{center}
\end{figure}

\noindent Using the parton model to describe the process and assuming $q_{\perp,cm} \ll q$, we are able to determine the structure functions that appear in Eq.$\,$(\ref{e:angdecomp}) in terms of TMD FFs.  Within this framework (see Fig.$\,$\ref{f:epluseminusPartonFact}) the cross section for the reaction reads \cite{Boer:1997mf, Boer:2008fr}
\begin{align}
4\frac{P_a^0P_b^0d\sigma_{em}} {d^3\vec{P}_ad^3\vec{P}_b} &= \frac{2\alpha_{em}^2N_c} {q^6}\, L^{\mu\nu}_{\gamma\gamma}\,\sum_{q}e_{q}^2\,\int\! dp_a^+d^{2}\vec{p}_{a\perp} dp_b^-d^2\vec{p}_{b\perp}\,\delta^{(2)}(\vec{p}_{a\perp}+\vec{p}_{b\perp}-\vec{q}_\perp)\nonumber\\
&\hspace{0.5cm}\times\, {\rm Tr}\left(\Delta^{h_a/q}\gamma_\mu \bar{\Delta}^{h_b/q} \gamma_\nu\right) + \left\{\Delta \leftrightarrow \bar{\Delta}\right\},\label{e:epluseminusPartonCS}
\end{align}
where $N_c = 3$ is the number of colors, $e_q$ is the quark charge in units of $e>0$, $L^{\mu\nu}_{\gamma\gamma}$ is given in (\ref{e:leptens2}), and $\left\{\Delta \leftrightarrow \bar{\Delta}\right\}$ takes into account the graph in Fig.$\,$\ref{f:epluseminusPartonFact}(b).  The summation in Eq.$\,$(\ref{e:epluseminusPartonCS}) then is only over quark flavors.  The arguments of the correlators have been suppressed but are implied as $\Delta^{h_a/q}(p_a;P_a,S_a|\bar{n})$ and $\bar{\Delta}^{h_b/q}(p_b;P_b,S_b|n)$.  Their operator definitions are given, respectively, by
\begin{align}
\Delta_{ij}^{h_a/q}(p_a;P_a;S_a|\bar{n}) = \sum_{X}\hspace{-0.55cm}\int\int \!\!\frac{d^4 \xi} {(2\pi)^4}e^{ip_a\cdot \xi}\langle 0|&\mathcal{W}_1(\infty,\xi|\bar{n})\psi^q_i(\xi)|P_a,S_a;X\rangle\nonumber\\[-0.4cm]
&\times\,\langle P_a,S_a;X|\bar{\psi}^q_j(0)\mathcal{W}_2(0,\infty|\bar{n})|0\rangle\,, \label{e:DeltaCorrel}
\end{align}
and
\begin{align}
\bar{\Delta}_{ij}^{h_b/q}(p_b;P_b,S_b|n) = \sum_{X}\hspace{-0.55cm}\int\int\!\! \frac{d^4 \xi} {(2\pi)^4}e^{ip_b\cdot \xi}\langle 0|&\mathcal{W}_1(\infty,\xi|n)\bar{\psi}^q_j(\xi)|P_b,S_b;X\rangle\nonumber\\[-0.45cm]
&\hspace{0.35cm}\times\,\langle P_b,S_b;X|\psi^q_i(0)\mathcal{W}_2(0,\infty|n)|0\rangle\,, \label{e:DeltabarCorrel}
\end{align}
where $\bar{n}$ ($n$) is a lightlike vector conjugate to the direction of $P_a$ ($P_b$), and $\mathcal{W}_1,\,\mathcal{W}_2$ are Wilson lines ensuring the color gauge invariance of the correlators.  

In this case we expand the cross section to twist-2, i.e., to order $(q_{\perp,cm}/q)^0$.  In particular this means $\cos\xi\rightarrow 1$ and $\sin\xi\rightarrow 0$ in Eqs.$\,$(\ref{e:lcm}),$\,$(\ref{e:lprimecm}) for the lepton momenta.  We can also carry out the integrations over $p_a^+$ and $p_b^-$.  
The TMD correlators\footnote{We mention that one must deal with important technical issues in order to properly define TMD correlators --- see, e.g., \cite{Collins:book, Cherednikov:2009wk, GarciaEchevarria:2011rb, Collins:2012uy, Echevarria:2012js} for recent treatments of these matters.  Such complications will not affect the results of this work.} that result can then be written in terms of twist-2 Dirac structures:
\begin{align}
\Delta^{h_a/q}(z_a, \vec{p}_{a\perp};P_a,S_a|\bar{n}) &= z_a\left(\Delta^{h_a/q[\gamma^-]}\,\gamma^+ - \Delta^{h_a/q[\gamma^-\gamma_5]}\,\gamma^+\gamma_5 + \Delta^{h_a/q[i\sigma^{i-}\gamma_5]}\,i\sigma^{i+}\gamma_5\right),\label{e:DeltaExp}\\[0.1cm]
\bar{\Delta}^{h_b/q}(z_b, \vec{p}_{b\perp};P_b,S_b|n) &= z_b\left(\bar{\Delta}^{h_b/q[\gamma^+]}\,\gamma^- - \bar{\Delta}^{h_b/q[\gamma^+\gamma_5]}\,\gamma^-\gamma_5 + \bar{\Delta}^{h_b/q[i\sigma^{j+}\gamma_5]}\,i\sigma^{j-}\gamma_5\right)\label{e:DeltabarExp},
\end{align}
where 
\begin{equation}
\Delta^{h_a/q[\Gamma]}(z_a,\vec{p}_{a\perp};P_a,S_a|\bar{n}) = \frac{1} {4z_a}\, {\rm Tr}\left[\Delta^{h_a/q}(z_a,\vec{p}_{a\perp};P_a,S_a|\bar{n}) \,\Gamma\right], \label{e:DeltaTr}
\end{equation}
and likewise for $\bar{\Delta}^{h_b/q[\Gamma]}(z_b,\vec{p}_{b\perp};P_b,S_b|n)$.  The variables $z_a$, $z_b$ are lightcone momentum fractions that satisfy $P_a^- = z_a\, p_a^-$ and $P_b^+ = z_b\, p_b^+$.  The correlator (\ref{e:DeltaTr}) gives eight twist-2 TMD FFs \cite{Boer:1997mf}:
\begin{eqnarray}
\Delta^{h_a/q[\gamma^-]} \!\!\!&=&\!\!\! D_1^{h_a/q}(z_a,z_a^2\vec{p}_{a\perp}^{\,2}) + \frac{\epsilon_{\perp}^{ij}p_{a\perp}^iS_{a\perp}^j} {M_a} \,D_{1T}^{\perp\,h_a/q}(z_a,z_a^2\vec{p}_{a\perp}^{\,2})\,, \label{e:DTMD}\\[0.3cm]
\Delta^{h_a/q[\gamma^-\gamma_5]} \!\!\!&=&\!\!\! \Lambda_a \,G_{1L}^{h_a/q}(z_a,z_a^2\vec{p}_{a\perp}^{\,2}) + \frac{\vec{p}_{a\perp}\cdot\vec{S}_{a\perp}} {M_a}\, G_{1T}^{h_a/q}(z_a,z_a^2\vec{p}_{a\perp}^{\,2})\,,\label{e:GTMD}\\[0.3cm]
\Delta^{h_a/q[i\sigma^{i-}\gamma_5]} \!\!\!&=&\!\!\! S_{a\perp}^i \,H_{1T}^{h_a/q}(z_a,z_a^2\vec{p}_{a\perp}^{\,2}) - \frac{\epsilon_\perp^{ij}p_{a\perp}^{\,j}} {M_a} \,H_1^{\perp\,h_a/q}(z_a,z_a^2\vec{p}_{a\perp}^{\,2}) \nonumber \\
&&\hspace{-0.25cm} +\, \frac{p_{a\perp}^i} {M_a} \left[\Lambda_a\, H_{1L}^{\perp\,h_a/q}(z_a,z_a^2\vec{p}_{a\perp}^{\,2}) + \frac{\vec{p}_{a\perp}\cdot\vec{S}_{a\perp}} {M_a}\,H_{1T}^{\perp\,h_a/q}(z_a,z_a^2\vec{p}_{a\perp}^{\,2})\right], \label{e:HTMD} 
\end{eqnarray}
where $\epsilon_\perp^{ij} \equiv \epsilon^{-+ij}$ and $\epsilon^{0123} = 1$.  Similar expressions hold for $\bar{\Delta}^{h_b/q[\Gamma]}$ with $\Gamma\hspace{-0.025cm}=\hspace{-0.025cm}\gamma^+,\,\gamma^+\gamma_5,\,i\sigma^{i+}\gamma_5$ if one keeps in mind the relation~\cite{Tangerman:1994eh}
\begin{eqnarray}
\bar{\Delta}^{h/q[\Gamma]} = 
	\begin{cases}
	+ \,\Delta^{h/\bar{q}[\Gamma]} \;{\rm for}\; \gamma^{\mu},\,i\sigma^{\mu\nu}\gamma_5 \\
	- \Delta^{h/\bar{q}[\Gamma]} \;{\rm for}\; \eins,\, i\gamma_5,\, \gamma^{\mu}\gamma_5\,, \label{e:qbarCorrelRel}
	\end{cases}
\end{eqnarray}  
and notes that $h_b$ in this situation has a large plus- (rather than a large minus-) component of momentum.  That is, one has 
\begin{eqnarray}
\bar{\Delta}^{h_b/q[\gamma^+]} \!\!\!&=&\!\!\! D_1^{h_b/q}(z_b,z_b^2\vec{p}_{b\perp}^{\,2}) - \frac{\epsilon_{\perp}^{ij}p_{b\perp}^iS_{b\perp}^j} {M_b} \,D_{1T}^{\perp\,h_b/q}(z_b,z_b^2\vec{p}_{b\perp}^{\,2})\,, \label{e:DTMDanti}\\[0.3cm]
\bar{\Delta}^{h_b/q[\gamma^+\gamma_5]} \!\!\!&=&\!\!\! -\Lambda_b \,G_{1L}^{h_b/q}(z_b,z_b^2\vec{p}_{b\perp}^{\,2}) - \frac{\vec{p}_{b\perp}\cdot\vec{S}_{b\perp}} {M_b}\, G_{1T}^{h_b/q}(z_b,z_b^2\vec{p}_{b\perp}^{\,2})\,,\label{e:GTMDanti}\\[0.3cm]
\bar{\Delta}^{h_b/q[i\sigma^{i+}\gamma_5]} \!\!\!&=&\!\!\! S_{b\perp}^i \,H_{1T}^{h_b/q}(z_b,z_b^2\vec{p}_{b\perp}^{\,2}) + \frac{\epsilon_\perp^{ij}p_{b\perp}^{\,j}} {M_b} \,H_1^{\perp\,h_b/q}(z_b,z_b^2\vec{p}_{b\perp}^{\,2}) \nonumber \\
&&\hspace{-0.25cm} +\, \frac{p_{b\perp}^i} {M_b} \left[\Lambda_b\, H_{1L}^{\perp\,h_b/q}(z_b,z_b^2\vec{p}_{b\perp}^{\,2}) + \frac{\vec{p}_{b\perp}\cdot\vec{S}_{b\perp}} {M_b}\,H_{1T}^{\perp\,h_b/q}(z_b,z_b^2\vec{p}_{b\perp}^{\,2})\right]. \label{e:HTMDanti} 
\end{eqnarray}

Substituting Eqs.$\,$(\ref{e:DeltaExp}),$\,$(\ref{e:DeltabarExp}) into (\ref{e:epluseminusPartonCS}) and contracting the Lorentz indices gives us
\begin{align}
4\frac{P_a^0P_b^0d\sigma_{em}} {d^3\vec{P}_ad^3\vec{P}_b} &= \frac{4\alpha_{em}^2N_c} {q^4}\,z_az_b\,\sum_{q}e_{q}^2\!\int\! d^{2}\vec{p}_{a\perp}d^2\vec{p}_{b\perp}\,\delta^{(2)}(\vec{p}_{a\perp}+\vec{p}_{b\perp}-\vec{q}_\perp)\nonumber\\
&\hspace{0.5cm}\times\, \left[(1+\cos^2\theta)\left(\Delta^{h_a/q[\gamma^-]}\bar{\Delta}^{h_b/q[\gamma^+]} + \Delta^{h_a/q[\gamma^-\gamma_5]}\bar{\Delta}^{h_b/q[\gamma^+\gamma_5]}\right)\right.\nonumber\\
&\hspace{1.25cm} +\,\sin^2\theta\Big(\cos2\phi\,(\delta^{i1}\delta^{j1}-\delta^{i2}\delta^{j2}) + \sin2\phi\,(\delta^{i1}\delta^{j2} + \delta^{i2}\delta^{j1})\Big)\nonumber\\
	&\hspace{6.9cm}\times\,\Delta^{h_a/q[i\sigma^{i-}\gamma_5]}\bar{\Delta}^{h_b/q[i\sigma^{j+}\gamma_5]} \nonumber\\
&\hspace{1.25cm} \left.+\,2\lambda_e\cos\theta\left(\Delta^{h_a/q[\gamma^-\gamma_5]}\bar{\Delta}^{h_b/q[\gamma^+]} + \Delta^{h_a/q[\gamma^-]}\bar{\Delta}^{h_b/q[\gamma^+\gamma_5]}\right)\right] \nonumber \\
&\hspace{0.5cm}+\, \left\{\Delta \leftrightarrow \bar{\Delta}\right\}.
\label{e:epluseminusPartonCS2}
\end{align}

We now explicitly write Eq.$\,$(\ref{e:epluseminusPartonCS2}) in terms of TMD FFs and compare the result to Eq.$\,$(\ref{e:angdecomp}) in order to obtain values for the structure functions at twist-2.  We define the convolution of TMD FFs in transverse momentum space as
\begin{align}
&\hspace{-0.25cm}C\!\left[w(\vec{p}_{a\perp},\vec{p}_{b\perp})D_1\bar{D}_2\right] \equiv 4z_az_bN_c\,\sum_q e_q^2\int\! d^2\vec{p}_{a\perp}d^2\vec{p}_{b\perp}\,\delta^{(2)}(\vec{p}_{a\perp}+\vec{p}_{b\perp}-\vec{q}_{\perp})\,w(\vec{p}_{a\perp},\vec{p}_{b\perp})\nonumber\\
&\hspace{0.5cm}\times\,\left[D_1^{h_a/q}(z_a,z_a^2\vec{p}_{a\perp}^{\,2})D_2^{h_b/\bar{q}}(z_b,z_b^2\vec{p}_{b\perp}^{\,2})+ \,D_1^{h_a/\bar{q}}(z_a,z_a^2\vec{p}_{a\perp}^{\,2})D_2^{h_b/q}(z_b,z_b^2\vec{p}_{b\perp}^{\,2})\right]
\end{align}
and introduce the following combinations of structure functions:
\begin{align}
&F_{TU}^{\sin(2\phi + \phi_a)} \equiv \frac{1} {2}\left(F_{TU}^{\cos2\phi} + F_{TU}^{\sin2\phi}\right), \hspace{0.35cm} F_{TU}^{\sin(2\phi - \phi_a)} \equiv \frac{1} {2}\left(F_{TU}^{\sin2\phi} - F_{TU}^{\cos2\phi}\right), \nonumber\\[0.2cm]
&F_{UT}^{\sin(2\phi + \phi_b)} \equiv \frac{1} {2}\left(F_{UT}^{\cos2\phi} + F_{UT}^{\sin2\phi}\right), \hspace{0.35cm} F_{UT}^{\sin(2\phi - \phi_b)} \equiv \frac{1} {2}\left(F_{UT}^{\sin2\phi}-F_{UT}^{\cos2\phi}\right), \nonumber\\[0.2cm]
&F_{LT}^{\cos(2\phi - \phi_b)} \equiv \frac{1} {2}\left(F_{LT}^{\cos2\phi} + F_{LT}^{\sin2\phi}\right), \hspace{0.35cm}  F_{LT}^{\cos(2\phi + \phi_b)} \equiv \frac{1} {2}\left(F_{LT}^{\cos2\phi} - F_{LT}^{\sin2\phi}\right), \nonumber\\[0.2cm]
&F_{TL}^{\cos(2\phi - \phi_a)} \equiv \frac{1} {2}\left(F_{TL}^{\cos2\phi} + F_{TL}^{\sin2\phi}\right), \hspace{0.35cm}  F_{TL}^{\cos(2\phi + \phi_a)} \equiv \frac{1} {2}\left(F_{TL}^{\cos2\phi} - F_{TL}^{\sin2\phi}\right), \nonumber\\[0.2cm]
&F_{TT}^{\cos(2\phi + \phi_a - \phi_b)} \equiv \frac{1} {2}\left(\bar{F}_{TT}^{\cos2\phi} - \bar{F}_{TT}^{\sin2\phi}\right),  \hspace{0.35cm} F_{TT}^{\cos(2\phi - \phi_a + \phi_b)} \equiv \frac{1} {2}\left(\bar{F}_{TT}^{\cos2\phi} + \bar{F}_{TT}^{\sin2\phi}\right), \nonumber\\[0.2cm]
&F_{TT}^{\cos(2\phi - \phi_a - \phi_b)} \equiv \frac{1} {2}\left(F_{TT}^{\cos2\phi} + F_{TT}^{\sin2\phi}\right),  \hspace{0.35cm} F_{TT}^{\cos(2\phi + \phi_a + \phi_b)} \equiv \frac{1} {2}\left(F_{TT}^{\cos2\phi} - F_{TT}^{\sin2\phi}\right). 
\end{align}
We also find it convenient to define the following weights:
\begin{align}
&w_{0} \equiv \frac{\vec{p}_{a\perp}^{\,2}}{2M_{a}^{2}}\,, \hspace{0.35cm} \bar{w}_{0} \equiv \frac{\vec{p}_{b\perp}^{\,2}}{2M_{b}^{2}}\,,\hspace{0.35cm}w_{0}^{\prime}  \equiv  \frac{\vec{p}_{a\perp}\cdot\vec{p}_{b\perp}}{M_{a}M_{b}}\,,  \hspace{0.35cm} w_{1}  \equiv  \frac{\hat{h}\cdot\vec{p}_{a\perp}}{M_{a}}\,,\hspace{0.35cm}\bar{w}_{1}  \equiv  \frac{\hat{h}\cdot\vec{p}_{b\perp}}{M_{b}}\,,\label{e:w0} \\[0.3cm]
&w_{2}  \equiv  \frac{2(\hat{h}\cdot \vec{p}_{a\perp})^2-\vec{p}_{a\perp}^{\,2}} {2M_a^2}\,,  \hspace{0.35cm}\bar{w}_{2}  \equiv \frac{2(\hat{h}\cdot \vec{p}_{b\perp})^2-\vec{p}_{b\perp}^{\,2}} {2M_b^2}\,,\hspace{0.35cm}w_{3}  \equiv  \frac{2(\hat{h}\cdot \vec{p}_{a\perp})(\hat{h}\cdot\vec{p}_{b\perp})-\vec{p}_{a\perp}\cdot\vec{p}_{b\perp}} {M_aM_b}\,,\\[0.3cm]
&w_{4}  \equiv  \frac{4(\hat{h}\cdot\vec{p}_{a\perp})^2(\hat{h}\cdot\vec{p}_{b\perp}) - 2(\hat{h}\cdot\vec{p}_{a\perp})(\vec{p}_{a\perp}\cdot\vec{p}_{b\perp}) - \vec{p}_{a\perp}^{\,2}(\hat{h}\cdot\vec{p}_{b\perp})} {2M_a^2 M_b}\,, \\[0.3cm] 
&\bar{w}_{4}  \equiv \frac{4(\hat{h}\cdot\vec{p}_{b\perp})^2(\hat{h}\cdot\vec{p}_{a\perp}) - 2(\hat{h}\cdot\vec{p}_{b\perp})(\vec{p}_{a\perp}\cdot\vec{p}_{b\perp}) - \vec{p}_{b\perp}^{\,2}(\hat{h}\cdot\vec{p}_{a\perp})} {2M_a M_b^2}\,,\\[0.3cm]
 &w_{4}^\prime  \equiv  \frac{\vec{p}_{a\perp}^{\,2}\vec{p}_{b\perp}^{\,2} - 2\vec{p}_{a\perp}^{\,2}(\hat{h}\cdot \vec{p}_{b\perp})^2 - 2(\hat{h}\cdot \vec{p}_{a\perp})^{\,2}\vec{p}_{b\perp}^{\,2} + 4(\hat{h}\cdot\vec{p}_{a\perp})(\hat{h}\cdot\vec{p}_{b\perp})[2(\hat{h}\cdot\vec{p}_{a\perp})(\hat{h}\cdot\vec{p}_{b\perp})-(\vec{p}_{a\perp}\cdot\vec{p}_{b\perp})]} {4M_a^2M_b^2}\,, \label{e:w4}
\end{align}
where $\hat{h} \equiv \vec{q}_{\perp,cm}/|\vec{q}_{\perp,cm}|$, and we have suppressed the arguments of the $w$'s for brevity.  Note also we define $H_1^{h/q}(z,z^2\vec{p}_{\perp}^{\,2})\equiv\left[H_{1T}^{h/q}(z,z^2\vec{p}_{\perp}^{\,2}) + (\vec{p}_{\perp}^{\,2}/2M^2)\,H_{1T}^{\perp\,h/q}(z,z^2\vec{p}_{\perp}^{\,2})\right]$.  In the end, one finds
\begin{align}
&F_{UU}^{1} = C\!\left[D_1\bar{D}_1\right], 
\hspace{0.3cm} F_{UU}^{\cos2\phi} = C\!\left[w_3\,H_{1}^{\perp}\bar{H}_{1}^{\perp}\right],
\hspace{0.3cm} F_{LU}^{\sin2\phi} = -C\!\left[w_3\,H_{1L}^{\perp}\bar{H}_{1}^{\perp}\right],
\hspace{0.3cm} F_{UL}^{\sin2\phi} = C\!\left[w_3\,H_{1}^{\perp}\bar{H}_{1L}^{\perp}\right], \label{e:F1}\\[0.5cm]
&F_{TU}^1 = C\!\left[w_1\,D_{1T}^{\perp}\bar{D}_1\right],
\hspace{0.35cm} F_{TU}^{\sin(2\phi+\phi_a)} \!=\! -C\!\left[w_4\,H_{1T}^\perp \bar{H}_1^\perp\right],
\hspace{0.35cm} F_{TU}^{\sin(2\phi-\phi_a)} = -C\!\left[\bar{w}_1H_1\bar{H}_{1}^\perp\right],\\[0.5cm]
&F_{UT}^1 = -C\!\left[\bar{w}_1\,D_{1}\bar{D}_{1T}^\perp\right],
\hspace{0.35cm} F_{UT}^{\sin(2\phi+\phi_b)} = C\!\left[\bar{w}_4\,H_{1}^\perp \bar{H}_{1T}^\perp\right],
\hspace{0.35cm} F_{UT}^{\sin(2\phi-\phi_b)} = C\!\left[w_1H_{1}^\perp\bar{H}_1\right],\\[0.5cm]
& F_{LL}^{1} = -C\!\left[G_{1L}\bar{G}_{1L}\right], 
\hspace{0.35cm} F_{LL}^{\cos2\phi} = C\!\left[w_3\,H_{1L}^{\perp}\bar{H}_{1L}^{\perp}\right],\\[0.5cm]
&F_{LT}^1 = -C\!\left[\bar{w}_1\,G_{1L}\bar{G}_{1T}\right],
\hspace{0.35cm} F_{LT}^{\cos(2\phi-\phi_b)} = C\!\left[w_1\,H_{1L}^\perp\bar{H}_1\right],
\hspace{0.35cm}F_{LT}^{\cos(2\phi+\phi_b)} = C\!\left[\bar{w}_4\,H_{1L}^\perp \bar{H}_{1T}^\perp\right], \\[0.5cm]
&F_{TL}^1 = -C\!\left[w_1\,G_{1T}\bar{G}_{1L}\right],\label{e:FTL1} 
\hspace{0.35cm} F_{TL}^{\cos(2\phi-\phi_a)} = C\!\left[\bar{w}_1\,H_1\bar{H}_{1L}^\perp\right],
\hspace{0.35cm} F_{TL}^{\cos(2\phi+\phi_a)} = C\!\left[w_4\,H_{1T}^\perp \bar{H}_{1L}^\perp\right]\\[0.5cm]
&F_{TT}^1 = C\!\left[\frac{w_3} {2}\left(D_{1T}^\perp\bar{D}_{1T}^\perp - G_{1T}\bar{G}_{1T}\right)\right],
\hspace{0.35cm} \bar{F}_{TT}^1 = -C\!\left[\frac{w_0^\prime} {2}\left(D_{1T}^\perp\bar{D}_{1T}^\perp + G_{1T}\bar{G}_{1T}\right)\right],\\[0.5cm]
&F_{TT}^{\cos(2\phi-\phi_a-\phi_b)} = C\!\left[H_1\bar{H}_1\right],
\hspace{0.35cm} F_{TT}^{\cos(2\phi+\phi_a-\phi_b)} = C\!\left[w_2H_{1T}^\perp\bar{H}_1\right], \\[0.5cm]
& F_{TT}^{\cos(2\phi-\phi_a+\phi_b)} = C\!\left[\bar{w}_2H_1\bar{H}_{1T}^\perp\right], 
\hspace{0.35cm} F_{TT}^{\cos(2\phi+\phi_a+\phi_b)} = C\!\left[w_4^\prime H_{1T}^\perp \bar{H}_{1T}^\perp\right], \\[0.5cm]
&G_{LU}^2 = -C\!\left[G_{1L}\bar{D}_1\right],
\hspace{0.35cm} G_{UL}^2 = C\!\left[D_{1}\bar{G}_{1L}\right],
\hspace{0.35cm} \bar{G}_{TU}^2 = -C\!\left[w_1\,G_{1T}\bar{D}_1\right],
\hspace{0.35cm} \bar{G}_{UT}^2 = C\!\left[\bar{w}_1\,D_1\bar{G}_{1T}\right],\\[0.5cm]
&\bar{G}_{LT}^2 = C\!\left[\bar{w}_1\,G_{1L}\bar{D}_{1T}^\perp\right],
\hspace{0.35cm} \bar{G}_{TL}^2 = C\!\left[w_1\,D_{1T}^\perp\bar{G}_{1L}\right],\\[0.5cm]
&\ddot{G}_{TT}^2 = C\!\left[\frac{w_3} {2} \,\left(D_{1T}^\perp\bar{G}_{1T} + G_{1T}\bar{D}_{1T}^\perp\right)\right],
\hspace{0.35cm} \hat{G}_{TT}^2 = C\!\left[\frac{w_0^\prime} {2} \,\left(D_{1T}^\perp\bar{G}_{1T} - G_{1T}\bar{D}_{1T}^\perp\right)\right]\,. \label{e:F11}
\end{align}
All the results (\ref{e:F1})--(\ref{e:F11}) agree with Ref.$\!$\cite{Boer:1997mf}.  Note, however, that in \cite{Boer:1997mf} different conventions for the azimuthal angles are used.

We now extend our discussion to the electroweak case where the electron and positron can annihilate into a $Z$-boson.  That is, we will calculate the last two terms in Eq.$\,$(\ref{e:epluseminusCross}) within the TMD parton model framework at twist-2.  We mention that a general (model-independent) description becomes much more involved in particular because one no longer has the parity constraint on the hadronic tensor, and, thus, more structure functions enter.  We will not pursue such an analysis here.  One can follow a similar procedure to the method outlined already in this section in order to write the cross section in terms of twist-2 TMD FFs.  For the case of unpolarized leptons, this gives us the result
\begin{align}
4&\frac{P_{a}^{0}P_{b}^{0}d\sigma_{ew}}{d^{3}\vec{P}_{a}\,d^{3}\vec{P}_{b}} =\nonumber\\[0.3cm]
&\hspace{0.6cm}\Big[(1+\cos^{2}\theta) F_{UU}^{1,ew}+(\cos\theta) F_{UU}^{2,ew} +(\sin^{2}\theta\cos2\phi) F_{UU}^{\cos2\phi,ew}+(\sin^{2}\theta\sin2\phi) F_{UU}^{\sin2\phi,ew}\Big] \nonumber\\ 
&\hspace{0.22cm} +\Lambda_{a}\,\Big[(1+\cos^{2}\theta) F_{LU}^{1,ew}+(\cos\theta) F_{LU}^{2,ew}+(\sin^{2}\theta\sin2\phi) F_{LU}^{\sin2\phi,ew}+(\sin^{2}\theta\cos2\phi) F_{LU}^{\cos2\phi,ew}\Big]\nonumber \\
&\hspace{0.22cm}+|\vec{S}_{a\perp}|\left[\sin\phi_{a}\left((1+\cos^{2}\theta) F_{TU}^{1,ew}+(\cos\theta) F_{TU}^{2,ew}\right)\right.\nonumber \\
& \hspace{1.6cm}+\cos\phi_{a}\left((1+\cos^{2}\theta) \bar{F}_{TU}^{1,ew}+(\cos\theta)\bar{F}_{TU}^{2,ew}\right)\nonumber \\
 & \hspace{1.6cm}+\sin(2\phi-\phi_{a})(\sin^{2}\theta) F_{TU}^{\sin(2\phi-\phi_{a})}+\cos(2\phi-\phi_{a})(\sin^{2}\theta) F_{TU}^{\cos(2\phi-\phi_{a})}\nonumber \\
 &  \hspace{1.6cm}\left.+\sin(2\phi+\phi_{a})(\sin^{2}\theta)F_{TU}^{\sin(2\phi+\phi_{a})}+\cos(2\phi+\phi_{a})(\sin^{2}\theta) F_{TU}^{\cos(2\phi+\phi_{a})}\right]\nonumber\\
&\hspace{0.22cm} +\Lambda_{b}\,\Big[(1+\cos^{2}\theta)F_{UL}^{1,ew}+(\cos\theta) F_{UL}^{2,ew}+(\sin^{2}\theta\,\sin2\phi) F_{UL}^{\sin2\phi,ew}+(\sin^{2}\theta\,\cos2\phi) F_{UL}^{\cos2\phi,ew}\Big]\nonumber \\
 &\hspace{0.22cm}+|\vec{S}_{b\perp}|\left[\sin\phi_{b}\left((1+\cos^{2}\theta) F_{UT}^{1,ew}+(\cos\theta) F_{UT}^{2,ew}\right)\right.\nonumber \\
& \hspace{1.6cm}+\cos\phi_{b}\left((1+\cos^{2}\theta)\bar{F}_{UT}^{1,ew}+(\cos\theta) \bar{F}_{UT}^{2,ew}\right)\nonumber \\
 & \hspace{1.6cm}+\sin(2\phi-\phi_{b}) (\sin^{2}\theta)F_{UT}^{\sin(2\phi-\phi_{b})}+\cos(2\phi-\phi_{b})(\sin^{2}\theta) F_{UT}^{\cos(2\phi-\phi_{b})}\nonumber \\
 &  \hspace{1.6cm}\left.+\sin(2\phi+\phi_{b})(\sin^{2}\theta)F_{UT}^{\sin(2\phi+\phi_{b})}+\cos(2\phi+\phi_{b})(\sin^{2}\theta)F_{UT}^{\cos(2\phi+\phi_{b})}\right]\nonumber\\
&\hspace{0.22cm}+ \Lambda_{a}\, \Lambda_{b}\,\Big[(1+\cos^{2}\theta)F_{LL}^{1,ew}+(\cos\theta) F_{LL}^{2,ew}+(\sin^{2}\theta\cos2\phi) F_{LL}^{\cos2\phi,ew}+(\sin^{2}\theta\sin2\phi) F_{LL}^{\sin2\phi,ew}\Big]\nonumber \\
&\hspace{0.22cm}+ \Lambda_{a}\,|\vec{S}_{b\perp}|\left[\cos\phi_{b}\left((1+\cos^{2}\theta) F_{LT}^{1,ew}+(\cos\theta) F_{LT}^{2,ew}\right)\right.\nonumber \\
 &\hspace{2.1cm} +\sin\phi_{b}\left((1+\cos^{2}\theta) \bar{F}_{LT}^{1,ew}+(\cos\theta) \bar{F}_{LT}^{2,ew}\right)\nonumber \\
 &\hspace{2.1cm} +\sin(2\phi-\phi_{b})(\sin^{2}\theta) F_{LT}^{\sin(2\phi-\phi_{b}),ew}+\cos(2\phi-\phi_{b})(\sin^{2}\theta) F_{LT}^{\cos(2\phi-\phi_{b}),ew}\nonumber \\
 &\hspace{2.1cm} \left.+\sin(2\phi+\phi_{b})(\sin^{2}\theta) F_{LT}^{\sin(2\phi+\phi_{b}),ew}+\cos(2\phi+\phi_{b})(\sin^{2}\theta) F_{LT}^{\cos(2\phi+\phi_{b}),ew}\right]\nonumber \\
 &\hspace{0.22cm}+ |\vec{S}_{a\perp}|\,\Lambda_{b}\left[\cos\phi_{a}\left((1+\cos^{2}\theta) F_{TL}^{1,ew}+(\cos\theta) F_{TL}^{2,ew}\right)\right.\nonumber \\
 &\hspace{2.1cm} +\sin\phi_{a}\left((1+\cos^{2}\theta)\bar{F}_{TL}^{1,ew}+(\cos\theta) \bar{F}_{TL}^{2,ew}\right)\nonumber \\
 &\hspace{2.1cm} +\sin(2\phi-\phi_{a})(\sin^{2}\theta) F_{TL}^{\sin(2\phi-\phi_{a}),ew}+\cos(2\phi-\phi_{a})(\sin^{2}\theta) F_{TL}^{\cos(2\phi-\phi_{a}),ew}\nonumber \\
 &\hspace{2.1cm} \left.+\sin(2\phi+\phi_{a})(\sin^{2}\theta) F_{TL}^{\sin(2\phi+\phi_{a}),ew}+\cos(2\phi+\phi_{a})(\sin^{2}\theta) F_{TL}^{\cos(2\phi+\phi_{a}),ew}\right]\nonumber\\
 &\hspace{0.22cm}+ |\vec{S}_{a\perp}|\,|\vec{S}_{b\perp}|\left[\cos(\phi_{a}+\phi_{b})\left((1+\cos^{2}\theta)F_{TT}^{1,ew}+(\cos\theta) F_{TT}^{2,ew}\right)\right.\nonumber \\
 &\hspace{2.5cm} +\cos(\phi_{a}-\phi_{b})\left((1+\cos^{2}\theta)\bar{F}_{TT}^{1,ew}+(\cos\theta) \bar{F}_{TT}^{2,ew}\right)\nonumber \\
 &\hspace{2.5cm} +\sin(\phi_{a}+\phi_{b})\left((1+\cos^{2}\theta)\ddot{F}_{TT}^{1,ew}+(\cos\theta) \ddot{F}_{TT}^{2,ew}\right)\nonumber \\
 & \hspace{2.5cm}+\sin(\phi_{a}-\phi_{b})\left((1+\cos^{2}\theta)\hat{F}_{TT}^{1,ew}+(\cos\theta) \hat{F}_{TT}^{2,ew}\right)\nonumber \\
 &\hspace{2.5cm} +\cos(2\phi-\phi_{a}-\phi_{b})(\sin^{2}\theta) F_{TT}^{\cos(2\phi-\phi_{a}-\phi_{b}),ew}\nonumber\\
 &\hspace{2.5cm}+\cos(2\phi-\phi_{a}+\phi_{b})(\sin^{2}\theta) F_{TT}^{\cos(2\phi-\phi_{a}+\phi_{b}),ew}\nonumber \\
 &\hspace{2.5cm}  +\cos(2\phi+\phi_{a}-\phi_{b})(\sin^{2}\theta) F_{TT}^{\cos(2\phi+\phi_{a}-\phi_{b}),ew}\nonumber\\
 &\hspace{2.5cm}+\cos(2\phi+\phi_{a}+\phi_{b})(\sin^{2}\theta) F_{TT}^{\cos(2\phi+\phi_{a}+\phi_{b}),ew}\nonumber \\
 &\hspace{2.5cm}  +\sin(2\phi-\phi_{a}-\phi_{b})(\sin^{2}\theta) F_{TT}^{\sin(2\phi-\phi_{a}-\phi_{b}),ew}\nonumber\\
 &\hspace{2.5cm}+\sin(2\phi-\phi_{a}+\phi_{b})(\sin^{2}\theta) F_{TT}^{\sin(2\phi-\phi_{a}+\phi_{b}),ew}\nonumber \\
 &\hspace{2.5cm} +\sin(2\phi+\phi_{a}-\phi_{b})(\sin^{2}\theta)F_{TT}^{\text{\ensuremath{\sin}}(2\phi+\phi_{a}-\phi_{b}),ew}\nonumber\\
 &\hspace{2.5cm}\left.+\sin(2\phi+\phi_{a}+\phi_{b})(\sin^{2}\theta) F_{TT}^{\sin(2\phi+\phi_{a}+\phi_{b}),ew}\right]. \label{e:angdecompEW}
\end{align}
We leave the actual expressions of the structure functions in terms of twist-2 TMD FFs for Appendix~\ref{a:ew} since they are quite a bit ``messier'' than the pure electromagnetic case given in Eqs.$\,$(\ref{e:F1})--(\ref{e:F11}).  We can also allow for lepton polarization, where it then becomes convenient to write the cross section as
\begin{equation}
4\frac{P_{a}^{0}P_{b}^{0}d\Delta\sigma_{ew}} {d^{3}\vec{P}_{a}\,d^{3}\vec{P}_{b}}\equiv\frac{1}{2}\left[\left(4\frac{P_{a}^{0}P_{b}^{0}d\sigma_{ew}^{\lambda_e=+1}}{d^{3}\vec{P}_{a}\,d^{3}\vec{P}_{b}}\right)-\left(4\frac{P_{a}^{0}P_{b}^{0}d\sigma_{ew}^{\lambda_e=-1}}{d^3\vec{P}_{a}\,d^{3}\vec{P}_{b}}\right)\right].\label{eq:DefPolLept}
\end{equation}
One finds this cross section has the exact same angular decomposition as Eq.$\,$(\ref{e:angdecompEW}).  Obviously the structure functions that enter Eq.$\,$(\ref{eq:DefPolLept}) have different values than those in (\ref{e:angdecompEW}), and we give these results in Appendix~\ref{a:ew} as well.  Thus, we have for the first time a complete framework for the study of TMD FFs within $e^+e^-\!\to h_ah_bX$ including electroweak terms and the polarization of all particles.  We also consider the reactions  $e^+e^-\! \rightarrow h\,jet\,X$ and $e^+e^-\! \rightarrow h\,X$ in Appendix~\ref{a:hjet}.  Note that the chiral-even TMD FFs in principle can also be studied by looking at a hadron inside a jet in $e^+e^-\!\rightarrow (h\,jet)\,X$~\cite{Yuan:2007nd}.

\begin{figure}[t]
\begin{center}
\includegraphics[width=11cm]{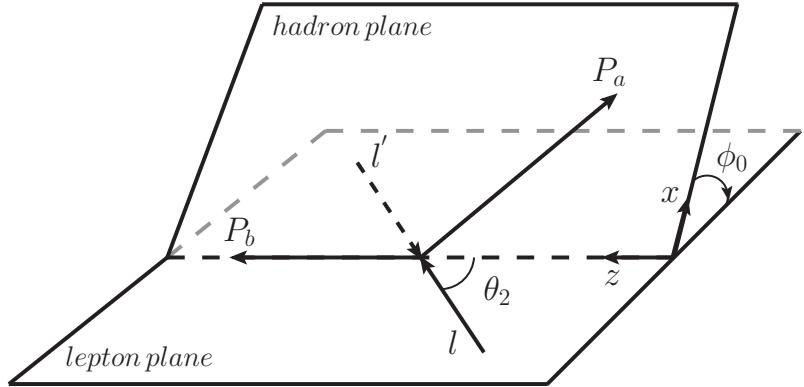}
\caption[]{Analogue of the Gottfried-Jackson frame for $e^+e^-\!\!\rightarrow h_a h_b\, X$.  The incoming electron makes an angle $\theta_2$ w.r.t.~the $+z$-axis defined by $\vec{P}_b$, and the hadron $h_a$ moves in a plane that forms an angle $\phi_0$ w.r.t.~the lepton plane.}
 \label{f:GJframe}
\end{center}
\end{figure}

We mention that the GJ frame differs from the CS frame by a rotation about the $y$-axis.  It turns out that the angle of this rotation is exactly the angle $\xi$ defined immediately after Eq.$\,$(\ref{e:cmCStrans}), which is zero to order $(q_{\perp,cm}/q)^0$.  Therefore, the frames are equivalent at leading twist.  Consequently, one can translate our twist-2 parton model results that involve CS angles to ones that involve GJ angles by making the replacements $\theta \to \theta_2$ and $\phi \to \phi_0$ in the angular prefactors and keeping the same values for the structure functions.  Here $\theta_2$ and $\phi_0$ are angles in the GJ frame (see Fig.$\,$\ref{f:GJframe}).  In this way our results have an exact connection to experimental analyses performed in the GJ frame and directly allow for the extraction of twist-2 TMD FFs.

%
%
%
%
\section{Discussion and conclusions}
\label{s:DandC} 

Several additional comments are in order on our results.  First, we note that if we make the identifications $(4P_a^0P_b^0d\sigma/d^3\vec{P}_ad^3\vec{P}_b)_{e^+e^-} \to (4l^0l^{\prime\,0}d\sigma/d^3\vec{l}\,d^3\vec{l}^{\,\prime})_{DY}$, $z_a\,(z_b)\to x_a \,(x_b)$, and $N_c\to 1/N_c$, the structure functions associated with unpolarized leptons (i.e., the $F$ ones) for the pure electromagnetic case are the same as those given in Ref.$\!$\cite{Arnold:2008kf} for Drell-Yan with the TMD PDFs replaced by their TMD FF analogues.  The only additional change one must remember is that $h_a$ ($h_b$) in the $e^+e^-$ case has a large minus- (plus-) component of momentum, whereas for Drell-Yan one normally uses the reverse convention.  This difference affects the TMD FF equivalents to the Sivers function $f_{1T}^\perp$ and Boer-Mulders function $h_1^\perp$ (i.e., the ``polarizing'' FF $D_{1T}^\perp$ and the Collins function $H_1^\perp$) since these have prefactors that contain $\epsilon_\perp^{ij}$.  The interchange of ``plus'' with ``minus'' between the TMD PDFs and TMD FFs introduces a negative sign when one replaces $f_{1T}^\perp$ with $D_{1T}^\perp$ and $h_1^\perp$ with $H_1^\perp$.  This adjustment is reflected in the values of the $F$ structure functions given above.  Along the same lines, one can easily transcribe our results in Appendix \ref{a:ew} to obtain the relevant expressions for Drell-Yan when one allows the $q\bar{q}$ pair to annihilate into a $Z$-boson.  Thus, we have for the first time full results for double-polarized Drell-Yan that include electroweak effects, which would be needed if such experiments were conducted at RHIC.

Next, 32 of the 72 total structure functions for the pure electromagnetic case are relevant at leading twist in that the angular prefactors associated with them enter into Eq.$\,$(\ref{e:epluseminusPartonCS2}).  Furthermore, note that 96 more angular structure functions show up at twist-2 once one allows for a virtual $Z$-boson.  For example, in the unpolarized case the two functions $F_{UU}^{2,ew}$ and $F_{UU}^{\sin2\phi,ew}$ arise due to $Z$-$Z$ and $\gamma$-$Z$ interference. In particular, the $\sin2\phi$ azimuthal dependence is generated by the imaginary part of the $Z$-propagator and is proportional to the decay width $\Gamma_{Z}$. This structure function leads to a double-Collins asymmetry\cite{Boer:2008fr}.  Also, notice that many of the FFs appear in more than one structure function.  Of these, $D_1$, which describes the fragmentation of an unpolarized quark into an unpolarized hadron, is the most accurately known \cite{Kretzer:2000yf, Bourhis:2000gs, Kniehl:2000fe, Kretzer:2001pz, Albino:2005me, Hirai:2007cx, deFlorian:2007aj, Albino:2008fy, Christova:2008te}.  Because certain FFs enter into multiple structure functions, measuring several of these angular modulations would provide important cross-checks on the formalism presented here, i.e., one must ascertain if a consistency exists between extractions of the same function through two different asymmetries.  Of course, to gain access to the relevant FFs would require the detection of (both longitudinally and transversely) polarized hadrons.  One interesting structure function to probe through, e.g., $e^+e^-\!\!\rightarrow\!\gamma^*\!\rightarrow\!\Lambda^\uparrow\pi\,X$ would be $F_{TU}^{\sin(2\phi-\phi_a)}$, which in principle would allow for the extraction of $H_1$.  Then one could access the transversity $h_1$ through the SIDIS reaction $ep^\uparrow\rightarrow e' \Lambda^\uparrow X$ in {\it collinear} factorization (i.e., $h_1(x)$ couples to $H_1(z)$).

In addition, there are a few measurements using polarized leptons that could be beneficial.  One useful measurement would be of $\bar{G}_{TU}^2$, which would give direct access to $G_{1T}$ without the need to first extract $G_{1L}$ as would be required if one only used unpolarized leptons (cf.~Eq.$\,$(\ref{e:FTL1})).  Furthermore, an example of a cross-check that could be performed would be to analyze $G_{LU}^2$ in order to obtain $G_{1L}$, and then compare that extraction to the one of $G_{1L}$ from $F_{LL}^1$ or from $F_{LL}^{2,ew}$.  The latter structure function ($F_{LL}^{2,ew}$) has not shown up in the literature so far.  Given that $F_{LL}^{2,ew}$ only has $Z$-$Z$ and $\gamma$-$Z$ contributions, one could access this function through a future ILC \cite{Baer:2013cma}.  Moreover, one can use the $\cos 2\phi$ double-Collins asymmetry that arises from $G_{UU}^{\cos 2\phi,ew}$ to access the Collins function.  This again is a new structure function that has not appeared in the literature before.  Such an experiment could again be performed at the ILC.  This result could then be checked against the Collins function that has been obtained recently from Belle and BABAR data \cite{Anselmino:2007fs}.  Given that the ILC would have around two orders of magnitude higher $cm$ energy than Belle and BABAR, such an analysis, as well as the aforementioned cross-check of $G_{1L}$, would also be an important test of the TMD evolution formalism and its application to phenomenology, which has been of recent interest \cite{Aybat:2011ge,  Echevarria:2012pw, Aybat:2011ta, Anselmino:2012aa, Bacchetta:2013pqa, Boer:2013zca, Sun:2013dya}.

Finally, the most studied of these structure functions from an experimental standpoint is $F_{UU}^{\cos2\phi}$, which is responsible for the azimuthal $\cos2\phi$ double-Collins asymmetry \cite{Boer:1997mf, Boer:2008fr}.  We repeat that this asymmetry has been measured by both the Belle Collaboration \cite{Abe:2005zx} and the BABAR Collaboration \cite{TheBABAR:2013yha} in order to obtain information on the Collins function $H_1^\perp$ \cite{Collins:1992kk}.  Along with an asymmetry involving the Collins function and the transversity $h_1$ that has been determined in SIDIS \cite{Airapetian:2004tw, Alexakhin:2005iw, Qian:2011py}, extractions of both functions have been performed \cite{Vogelsang:2005cs, Efremov:2006qm, Anselmino:2007fs}.  We mention again that one can also have $\cos 2\phi$ as well as a $\sin 2\phi$ double-Collins asymmetries that result from $Z$-$Z$ and $\gamma$-$Z$ reactions (for both unpolarized and polarized leptons), which would be beneficial to explore, although the $\sin 2\phi$ asymmetry is most likely numerically small because it only involves $\gamma$-$Z$ terms \cite{Boer:2008fr}.  Nevertheless, knowledge of $\gamma$-$Z$ interference terms in general would be needed for precision measurements of TMD FFs at Belle and BABAR.  Another T-odd FF similar to the Collins function is $D_{1T}^\perp$, which describes the fragmentation of an unpolarized quark into a transversely polarized hadron and becomes relevant in the detection of $\Lambda$'s --- see, e.g.,\cite{Anselmino:2000vs} and references therein.  The universality of both functions has also been a topic of interest \cite{Metz:2002iz, Collins:2004nx, Gamberg:2008yt, Yuan:2007nd, Yuan:2008yv, Meissner:2008yf, Boer:2010ya}.

To conclude, we have analyzed the production of almost back-to-back hadron pairs from electron-positron annihilation, allowing for the polarization of all particles involved.  We have given a general (model-independent) analysis for the pure electromagnetic case $e^+e^-\!\!\rightarrow \!\gamma^*\!\rightarrow\! h_ah_bX$ and also calculated, using the parton model, the relevant structure functions in terms of twist-2 TMD FFs.  Furthermore, we have studied the electroweak reaction $e^+e^-\!\!\rightarrow\! Z^*\!\rightarrow\! h_ah_bX$ (including $\gamma$-$Z$ interference) within this model.  This is the first time a complete framework has been presented for the examination of TMD FFs within $e^+e^-\!\to h_ah_bX$.  Note that the general form of the hadronic tensor found in Sect.$\,$\ref{s:DecompHad} can be readily used in triple-polarized SIDIS and for di-hadron fragmentation.  We have also discussed the importance of our results for future $e^+e^-$ experiments (especially ones with polarized leptons), which include cross-checks of TMD FF extractions and tests of TMD evolution.  Both of these applications involve structure functions that have not appeared in the literature before.  Moreover, we have given an explicit prescription of how our work can be translated to the Drell-Yan reaction.  This again is the first time full results are available for double-polarized Drell-Yan that include electroweak effects.  Such experiments could be performed at RHIC. 
\\[0.5cm]
%
%
\noindent
{\bf Acknowledgments:} $\!\!$The work of A.M. and D.P. has been supported by the NSF under Grant No.~PHY-1205942.  D.P. also acknowledges support from the RIKEN BNL Research Center.

%
%
%
%
\begin{appendices}
\appendixpage
\section{Electroweak twist-2 structure functions for polarized hadron pairs} \label{a:ew}
In this Appendix we give the values of the twist-2 electroweak structure functions that appear in Eqs.$\,$(\ref{e:angdecompEW}), (\ref{eq:DefPolLept}).  Recall that (\ref{eq:DefPolLept}) (for lepton polarization) has the exact same angular decomposition as (\ref{e:angdecompEW}), and we will use $G$ to denote the structure functions that enter into the former.  The same weights in Eqs.$\,$(\ref{e:w0})--(\ref{e:w4}) will enter into the electroweak case as well.  The relevant convolutions of TMD FFs in transverse momentum space read
\begin{align}
&\hspace{-0.25cm}C^q_{ew}\!\!\left[w(\vec{p}_{a\perp},\vec{p}_{b\perp})D_1\bar{D}_2\right] \equiv 4z_az_bN_c\,\int\! d^2\vec{p}_{a\perp}d^2\vec{p}_{b\perp}\,\delta^{(2)}(\vec{p}_{a\perp}+\vec{p}_{b\perp}-\vec{q}_{\perp})\,w(\vec{p}_{a\perp},\vec{p}_{b\perp})\nonumber\\
&\hspace{0.5cm}\times\,\left[D_1^{h_a/q}(z_a,z_a^2\vec{p}_{a\perp}^{\,2})D_2^{h_b/\bar{q}}(z_b,z_b^2\vec{p}_{b\perp}^{\,2})+ \,D_1^{h_a/\bar{q}}(z_a,z_a^2\vec{p}_{a\perp}^{\,2})D_2^{h_b/q}(z_b,z_b^2\vec{p}_{b\perp}^{\,2})\right],\\[0.3cm]
&\hspace{-0.25cm}\tilde{C}^q_{ew}\!\!\left[w(\vec{p}_{a\perp},\vec{p}_{b\perp})D_1\bar{D}_2\right] \equiv 4z_az_bN_c\,\int\! d^2\vec{p}_{a\perp}d^2\vec{p}_{b\perp}\,\delta^{(2)}(\vec{p}_{a\perp}+\vec{p}_{b\perp}-\vec{q}_{\perp})\,w(\vec{p}_{a\perp},\vec{p}_{b\perp})\nonumber\\
&\hspace{0.5cm}\times\,\left[D_1^{h_a/q}(z_a,z_a^2\vec{p}_{a\perp}^{\,2})D_2^{h_b/\bar{q}}(z_b,z_b^2\vec{p}_{b\perp}^{\,2})- \,D_1^{h_a/\bar{q}}(z_a,z_a^2\vec{p}_{a\perp}^{\,2})D_2^{h_b/q}(z_b,z_b^2\vec{p}_{b\perp}^{\,2})\right].
\end{align}
The ``symmetric'' convolution $C^q_{ew}$ is analogous to the one defined in the pure electromagnetic case while the ``antisymmetric'' convolution $\tilde{C}^q_{ew}$ is a new feature of the electroweak reaction.  The latter enters because the contribution from Fig.$\,$\ref{f:epluseminusPartonFact}(b) for terms that involve the axial part of the quark-$Z$ coupling differs in sign from Fig.$\,$\ref{f:epluseminusPartonFact}(a).  The weak charges are given by
\begin{eqnarray}
a_{q}=\left\{\!\! \begin{array}{c}
1-\tfrac{8}{3}\sin^{2}\theta_{W},\,q=u,\, c,\, t,\,\bar{u},\,\bar{c},\,\bar{t}\\
1+\tfrac{4}{3}\sin^{2}\theta_{W},\,q=d,\, s,\, b,\,\bar{d},\,\bar{s},\,\bar{b}
\end{array}\right. & ,& b_{q}=\left\{\! \!\begin{array}{c}
-1,\,q=u,\, c,\, t,\,\bar{u},\,\bar{c},\,\bar{t}\\
+1,\,q=d,\, s,\, b,\,\bar{d},\,\bar{s},\,\bar{b}
\end{array}\right..\label{eq:wcharge}\
\end{eqnarray}
We also find it convenient to use a shorthand notation for the following functions of $s=q^2$:
\begin{align}
\mathcal{F}_{\gamma\gamma}(s) & =  \frac{\alpha_{em}^{2}}{s^{2}}\,,\nonumber \\
\mathcal{F}_{ZZ}(s) & =  \frac{G_{F}^{2}M_{Z}^{4}\,}{128\pi^{2}\left((s-M_{Z}^{2})^{2}+\Gamma_{Z}^{2}M_{Z}^{2}\right)}\,,\nonumber \\
\mathcal{F}_{\gamma Z}(s) & =  \frac{\alpha_{em}\, G_{F}M_{Z}^{2}\,(s-M_{Z}^{2})}{4\sqrt{2}\,\pi\, s\,\left((s-M_{Z}^{2})^{2}+\Gamma_{Z}^{2}M_{Z}^{2}\right)}\,,\nonumber \\
\tilde{\mathcal{F}}_{\gamma Z}(s) & =  \frac{\alpha_{em}\, G_{F}M_{Z}^{2}\,\Gamma_{Z}M_{Z}}{4\sqrt{2}\,\pi\, s\,\left((s-M_{Z}^{2})^{2}+\Gamma_{Z}^{2}M_{Z}^{2}\right)}\,.\label{eq:ewFormFactors}
\end{align}
These functions show up in eight flavor-dependent combinations:
\begin{align}
\mathcal{F}_1^{q\pm}(s) &\equiv e_{q}^{2}\mathcal{F}_{\gamma\gamma}(s)+(1+a_{Z}^{2})(a_{q}^{2}\pm b_{q}^{2})\,\mathcal{F}_{ZZ}(s)-a_{Z} e_{q}a_{q}\mathcal{F}_{\gamma Z}(s)\,,\\[0.3cm]
\mathcal{F}_2^q(s) &\equiv 4a_{Z}a_{q}b_{q}\mathcal{F}_{ZZ}(s)-e_{q}b_{q}\mathcal{F}_{\gamma Z}(s)\,,\\[0.3cm]
\mathcal{F}_3^{q,a_Z}(s) &\equiv a_{Z}e_{q}b_{q}\tilde{\mathcal{F}}_{\gamma Z}(s)\,, \hspace{0.35cm} \mathcal{F}_3^{q}(s) \equiv e_{q}b_{q}\tilde{\mathcal{F}}_{\gamma Z}(s)\\[0.3cm]
\mathcal{F}_4^q(s) &\equiv 2(1+a_{Z}^{2}) a_{q}b_{q}\mathcal{F}_{ZZ}(s)-a_{Z}e_{q}b_{q}\mathcal{F}_{\gamma Z}(s)\,, \\[0.3cm]
\mathcal{F}_5^{q\pm}(s) &\equiv 2a_{Z}(a_{q}^{2}\pm b_{q}^{2})\,\mathcal{F}_{ZZ}(s)-e_{q}a_{q}\,\mathcal{F}_{\gamma Z}(s)\,.
\end{align}

Finally we are able to write down the twist-2 electroweak structure functions.  Note that the pure electromagnetic results are included again for completeness.  For the case of unpolarized leptons (Eq.$\,$(\ref{e:angdecompEW})) we have
\begin{align}
&F_{UU}^{1,ew}  =  \sum_{q}\mathcal{F}_1^{q+}(s)\,C^{q}_{ew}\!\left[D_{1}\, \bar{D}_{1}\right], 
\hspace{0.35cm}F_{UU}^{2,ew}  =  2\sum_{q} \mathcal{F}_2^{q}(s)\,\tilde{C}^{q}_{ew}\!\left[D_{1}\, \bar{D}_{1}\right], \label{e:Few1}\\[0.5cm]
&F_{UU}^{\cos2\phi,ew}  =  \sum_{q}\mathcal{F}_1^{q-}(s)\,C^{q}_{ew}\!\left[w_{3}\, H_{1}^{\perp}\, \bar{H}_{1}^{\perp}\right]\,,
\hspace{0.35cm}F_{UU}^{\sin2\phi,ew}  =  \sum_{q}\mathcal{F}_3^{q,a_z}(s)\,\tilde{C}^{q}_{ew}\!\left[w_{3}\, H_{1}^{\perp}\, \bar{H}_{1}^{\perp}\right],\\[0.5cm]
&F_{LU}^{1,ew}  = \sum_{q}\mathcal{F}_4^{q}(s)\,\tilde{C}^{q}_{ew}\!\left[G_{1L}\, \bar{D}_{1}\right], 
\hspace{0.35cm}F_{LU}^{2,ew}  = 2\sum_{q}\mathcal{F}_5^{q+}(s)\,C^{q}_{ew}\!\left[G_{1L}\, \bar{D}_{1}\right], \\[0.5cm]
&F_{LU}^{\sin2\phi,ew}  = -\sum_{q}\mathcal{F}_1^{q-}(s)\,C^{q}_{ew}\!\left[w_{3}\, H_{1L}^{\perp}\, \bar{H}_{1}^{\perp}\right],
\hspace{0.35cm}F_{LU}^{\cos2\phi,ew}  = \sum_{q}\mathcal{F}_3^{q,a_z}(s)\,\tilde{C}^{q}_{ew}\!\left[w_{3}\, H_{1L}^{\perp}\,\bar{H}_{1}^{\perp}\right],\\[0.5cm]
&F_{TU}^{1,ew}  = \sum_{q}\mathcal{F}_1^{q+}(s)\,C^{q}_{ew}\!\left[w_{1}\, D_{1T}^{\perp}\, \bar{D}_{1}\right],
\hspace{0.35cm}F_{TU}^{2,ew}  =  2\sum_{q}\mathcal{F}_2^{q}(s)\,\tilde{C}^{q}_{ew}\!\left[w_{1}\, D_{1T}^{\perp}\, \bar{D}_{1}\right],\\[0.5cm]
& \bar{F}_{TU}^{1,ew}  = \sum_{q}\mathcal{F}_4^{q}(s)\,\tilde{C}^{q}_{ew}\!\left[w_{1}\, G_{1T}\, \bar{D}_{1}\right],
\hspace{0.35cm}\bar{F}_{TU}^{2,ew}  = 2\sum_{q}\mathcal{F}_5^{q+}(s)\,C^{q}_{ew}\!\left[w_{1}\, G_{1T}\, \bar{D}_{1}\right],\\[0.5cm]
&F_{TU}^{\sin(2\phi-\phi_{a}),EW}  = -\sum_{q}\mathcal{F}_1^{q-}(s)\,C^{q}_{ew}\!\left[\bar{w}_{1}\, H_{1}\, \bar{H}_{1}^{\perp}\right],
\hspace{0.35cm}F_{TU}^{\cos(2\phi-\phi_{a}),ew}  =  \sum_{q}\mathcal{F}_3^{q,a_z}(s)\,\tilde{C}^{q}_{ew}\!\left[\bar{w}_{1}\, H_{1}\, \bar{H}_{1}^{\perp}\right],\\[0.5cm]
&F_{TU}^{\sin(2\phi+\phi_{a}),ew}  =  -\sum_{q}\mathcal{F}_1^{q-}(s)\,C^{q}_{ew}\!\left[w_{4}\, H_{1T}^{\perp}\, \bar{H}_{1}^{\perp}\right],
\hspace{0.35cm} F_{TU}^{\cos(2\phi+\phi_{a}),ew}  =  \sum_{q}\mathcal{F}_3^{q,a_z}(s)\,\tilde{C}^{q}_{ew}\!\left[w_{4}\, H_{1T}^{\perp}\, \bar{H}_{1}^{\perp}\right],\\[0.5cm]
&F_{UL}^{1,ew}  = -\sum_{q}\mathcal{F}_4^{q}(s)\,\tilde{C}^{q}_{ew}\!\left[D_{1}\, \bar{G}_{1L}\right],
\hspace{0.35cm}F_{UL}^{2,ew}  = -2\sum_{q}\mathcal{F}_5^{q+}(s)\,C^{q}_{ew}\!\left[D_{1}\, \bar{G}_{1L}\right], \\[0.5cm]
&F_{UL}^{\sin2\phi,ew}  = \sum_{q}\mathcal{F}_1^{q-}(s)\,C^{q}_{ew}\!\left[w_{3}\, H_{1}^{\perp}\, \bar{H}_{1L}^{\perp}\right],
\hspace{0.35cm}F_{UL}^{\cos2\phi,ew}  = -\sum_{q}\mathcal{F}_3^{q,a_z}(s)\,\tilde{C}^{q}_{ew}\!\left[w_{3}\, H_{1}^{\perp}\, \bar{H}_{1L}^{\perp}\right],\\[0.5cm]
&F_{UT}^{1,ew}  =  -\sum_{q}\mathcal{F}_1^{q+}(s)\,C^{q}_{ew}\!\left[\bar{w}_{1}\, D_{1}\, \bar{D}_{1T}^{\perp}\right],
\hspace{0.35cm}F_{UT}^{2,ew}  = -2\sum_{q}\mathcal{F}_2^{q}(s)\,\tilde{C}^{q}_{ew}\!\left[\bar{w}_{1}\, D_{1}\, \bar{D}_{1T}^{\perp}\right],\\[0.5cm]
&\bar{F}_{UT}^{1,ew}  =  -\sum_{q}\mathcal{F}_4^{q}(s)\,\tilde{C}^{q}_{ew}\!\left[\bar{w}_{1}\, D_{1}\, \bar{G}_{1T}\right],
\hspace{0.35cm}\bar{F}_{UT}^{2,ew}  = -2\sum_{q}\mathcal{F}_5^{q+}(s)\,C^{q}_{ew}\!\left[\bar{w}_{1}\, D_{1}\, \bar{G}_{1T}\right],\\[0.5cm]
&F_{UT}^{\sin(2\phi-\phi_{b}),ew}  =  \sum_{q}\mathcal{F}_1^{q-}(s)C^{q}_{ew}\!\left[w_{1}\, H_{1}^{\perp}\, \bar{H}_{1}\right],
\hspace{0.35cm}F_{UT}^{\cos(2\phi-\phi_{b}),ew}  =  -\sum_{q}\mathcal{F}_3^{q,a_z}(s)\,\tilde{C}^{q}_{ew}\!\left[w_{1}\, H_{1}^{\perp}\, \bar{H}_{1}\right],\\[0.5cm]
&F_{UT}^{\sin(2\phi+\phi_{b}),ew}  =  \sum_{q}\mathcal{F}_1^{q-}(s)\,C^{q}_{ew}\!\left[\bar{w}_{4}\, H_{1}^{\perp}\, \bar{H}_{1T}^{\perp}\right],
\hspace{0.35cm}F_{UT}^{\cos(2\phi+\phi_{b}),ew}  =  -\sum_{q}\mathcal{F}_3^{q,a_z}(s)\,\tilde{C}^{q}_{ew}\!\left[\bar{w}_{4}\, H_{1}^{\perp}\, \bar{H}_{1T}^{\perp}\right],\\[0.5cm]
&F_{LL}^{1,ew}  =  -\sum_{q}\mathcal{F}_1^{q+}(s)\,C^{q}_{ew}\!\left[G_{1L}\, \bar{G}_{1L}\right],
\hspace{0.35cm}F_{LL}^{2,ew}  =  -2\sum_{q}\mathcal{F}_2^{q}(s)\,\tilde{C}^{q}_{ew}\!\left[G_{1L}\, \bar{G}_{1L}\right], \\[0.5cm]
&F_{LL}^{\cos2\phi,ew}  =  \sum_{q}\mathcal{F}_1^{q-}(s)C^{q}_{ew}\!\left[w_{3}\, H_{1L}^{\perp}\, \bar{H}_{1L}^{\perp}\right],
\hspace{0.35cm}F_{LL}^{\sin2\phi,ew}  =  \sum_{q}\mathcal{F}_3^{q,a_z}(s)\,\tilde{C}^{q}_{ew}\!\left[w_{3}\, H_{1L}^{\perp}\, \bar{H}_{1L}^{\perp}\right],\\[0.5cm]
&F_{LT}^{1,ew}  = -\sum_{q}\mathcal{F}_1^{q+}(s)\,C^{q}_{ew}\!\left[\bar{w}_{1}\, G_{1L}\, \bar{G}_{1T}\right],
\hspace{0.35cm}F_{LT}^{2,ew} =  -2\sum_{q}\mathcal{F}_2^{q}(s)\,\tilde{C}^{q}_{ew}\!\left[\bar{w}_{1}\, G_{1L}\, \bar{G}_{1T}\right],\\[0.5cm]
&\bar{F}_{LT}^{1,ew}  = - \sum_{q}\mathcal{F}_4^{q}(s)\,\tilde{C}^{q}_{ew}\!\left[\bar{w}_{1}\, G_{1L}\, \bar{D}_{1T}^{\perp}\right],
\hspace{0.35cm}\bar{F}_{LT}^{2,ew}  =  2\sum_{q}\mathcal{F}_5^{q+}(s)\,C^{q}_{ew}\!\left[\bar{w}_{1}\, G_{1L}\, \bar{D}_{1T}^{\perp}\right],\\[0.5cm]
&F_{LT}^{\sin(2\phi-\phi_{b}),ew}  =  \sum_{q}\mathcal{F}_3^{q,a_z}(s)\,\tilde{C}^{q}_{ew}\!\left[w_{1}\, H_{1L}^{\perp}\, \bar{H}_{1}\right],
\hspace{0.35cm}F_{LT}^{\cos(2\phi-\phi_{b}),ew}  =  \sum_{q}\mathcal{F}_1^{q-}(s)\,C^{q}_{ew}\!\left[w_{1}\, H_{1L}^{\perp}\, \bar{H}_{1}\right],\\[0.5cm]
&F_{LT}^{\sin(2\phi+\phi_{b}),ew}  =  \sum_{q}\mathcal{F}_3^{q,a_z}(s)\,\tilde{C}^{q}_{ew}\!\left[\bar{w}_{4}\, H_{1L}^{\perp}\, \bar{H}_{1T}^{\perp}\right],
\hspace{0.35cm}F_{LT}^{\cos(2\phi+\phi_{b}),ew}  =  \sum_{q}\mathcal{F}_1^{q-}(s)C^{q}_{ew}\!\left[\bar{w}_{4}\, H_{1L}^{\perp}\, \bar{H}_{1T}^{\perp}\right],\\[0.5cm]
&F_{TL}^{1,ew}  =  -\sum_{q}\mathcal{F}_1^{q+}(s)\,C^{q}_{ew}\!\left[w_{1}\, G_{1T}\, \bar{G}_{1L}\right],
\hspace{0.35cm}F_{TL}^{2,ew}  =  -2\sum_{q}\mathcal{F}_2^{q}(s)\,\tilde{C}^{q}_{ew}\!\left[w_{1}\, G_{1T}\, \bar{G}_{1L}\right],\\[0.5cm]
&\bar{F}_{TL}^{1,ew}  =  -\sum_{q}\mathcal{F}_4^{q}(s)\,\tilde{C}^{q}_{ew}\!\left[w_{1}\, D_{1T}^{\perp}\, \bar{G}_{1L}\right],
\hspace{0.35cm}\bar{F}_{TL}^{2,ew}  =  2\sum_{q}\mathcal{F}_5^{q+}(s)\,C^{q}_{ew}\!\left[w_{1}\, D_{1T}^{\perp}\, \bar{G}_{1L}\right],\\[0.5cm]
&F_{TL}^{\sin(2\phi-\phi_{a}),ew}  =  \sum_{q}\mathcal{F}_3^{q,a_z}(s)\,\tilde{C}^{q}_{ew}\!\left[\bar{w}_{1}\, H_{1}\, \bar{H}_{1L}^{\perp}\right],
\hspace{0.35cm}F_{TL}^{\cos(2\phi-\phi_{a}),ew}  =  \sum_{q}\mathcal{F}_1^{q+}(s)\,C^{q}_{ew}\!\left[\bar{w}_{1}\, H_{1}\, \bar{H}_{1L}^{\perp}\right],\\[0.5cm]
&F_{TL}^{\sin(2\phi+\phi_{a}),ew}  =  \sum_{q}\mathcal{F}_3^{q,a_z}(s)\,\tilde{C}^{q}_{ew}\!\left[w_{4}\, H_{1T}^{\perp}\, \bar{H}_{1L}^{\perp}\right],
\hspace{0.35cm}F_{TL}^{\cos(2\phi+\phi_{a}),ew}  =  \sum_{q}\mathcal{F}_1^{q-}(s)\,C^{q}_{ew}\!\left[w_{4}\, H_{1T}^{\perp}\, \bar{H}_{1L}^{\perp}\right],\\[0.5cm]
&F_{TT}^{1,ew}  =  \sum_{q}\mathcal{F}_1^{q+}(s)\,C^{q}_{ew}\!\left[\frac{w_{3}} {2}\,(D_{1T}^{\perp}\, \bar{D}_{1T}^{\perp}-G_{1T}\, \bar{G}_{1T})\right],\\[0.5cm]
&F_{TT}^{2,ew}  = -\sum_{q}\mathcal{F}_2^{q}(s)\,\tilde{C}^{q}_{ew}\!\left[w_{3}\,(G_{1T}\, \bar{G}_{1T}-D_{1T}^{\perp}\, \bar{D}_{1T}^{\perp})\right],\\[0.5cm]
&\bar{F}_{TT}^{1,ew}  = - \sum_{q}\mathcal{F}_1^{q+}(s)\,C^{q}_{ew}\!\left[\frac{w_{0}^{\prime}} {2}\,(D_{1T}^{\perp}\, \bar{D}_{1T}^{\perp}+G_{1T}\, \bar{G}_{1T})\right],\\[0.5cm]
&\bar{F}_{TT}^{2,ew}  =  -\sum_{q}\mathcal{F}_2^{q}(s)\,\tilde{C}^{q}_{ew}\!\left[w_{0}^{\prime}\,(G_{1T}\, \bar{G}_{1T}+D_{1T}^{\perp}\, \bar{D}_{1T}^{\perp})\right],\\[0.5cm]
&\ddot{F}_{TT}^{1,ew}  =  -\sum_{q}\mathcal{F}_4^{q}(s)\,\tilde{C}^{q}_{ew}\!\left[\frac{w_{3}} {2}\,(D_{1T}^{\perp}\, \bar{G}_{1T}+G_{1T}\, \bar{D}_{1T}^{\perp})\right],\\[0.5cm]
&\ddot{F}_{TT}^{2,ew}  =  -\sum_{q}\mathcal{F}_5^{q+}(s)\,C^{q}_{ew}\!\left[w_{3}\,(D_{1T}^{\perp}\, \bar{G}_{1T}+G_{1T}\, \bar{D}_{1T}^{\perp})\right],\\[0.5cm]
&\hat{F}_{TT}^{1,ew}  =  -\sum_{q}\mathcal{F}_4^{q}(s)\,\tilde{C}^{q}_{ew}\!\left[\frac{w_{0}^{\prime}} {2}\,(D_{1T}^{\perp}\, \bar{G}_{1T}-G_{1T}\, \bar{D}_{1T}^{\perp})\right],\\[0.5cm]
&\hat{F}_{TT}^{2,ew}  =  -\sum_{q}\mathcal{F}_5^{q+}(s)\,C^{q}_{ew}\!\left[w_{0}^{\prime}\,(D_{1T}^{\perp}\, \bar{G}_{1T}-G_{1T}\, \bar{D}_{1T}^{\perp})\right],\\[0.5cm]
&F_{TT}^{\cos(2\phi-\phi_{a}-\phi_{b}),ew}  =  \sum_{q}\mathcal{F}_1^{q-}(s)\,C^{q}_{ew}\!\left[H_{1}\, \bar{H}_{1}\right],
\hspace{0.35cm}F_{TT}^{\cos(2\phi-\phi_{a}+\phi_{b}),ew}  =  \sum_{q}\mathcal{F}_1^{q-}(s)\,C^{q}_{ew}\!\left[\bar{w}_{2}\, H_{1}\, \bar{H}_{1T}^{\perp}\right],\\[0.5cm]
&F_{TT}^{\cos(2\phi+\phi_{a}-\phi_{b}),ew}  = \sum_{q}\mathcal{F}_1^{q-}(s)\,C^{q}_{ew}\!\left[w_{2}\, H_{1T}^{\perp}\, \bar{H}_{1}\right],
\hspace{0.3cm}F_{TT}^{\cos(2\phi+\phi_{a}+\phi_{b}),ew}  \!=\!  \sum_{q}\mathcal{F}_1^{q-}(s)\,C^{q}_{ew}\!\left[w_{4}^\prime\, H_{1T}^{\perp}\, \bar{H}_{1T}^{\perp}\right],\\[0.5cm]
&F_{TT}^{\sin(2\phi-\phi_{a}-\phi_{b}),ew}  =  \sum_{q}\mathcal{F}_3^{q,a_z}(s)\,\tilde{C}^{q}_{ew}\!\left[H_{1}\, \bar{H}_{1}\right],
\hspace{0.35cm}F_{TT}^{\sin(2\phi-\phi_{a}+\phi_{b}),ew}  = \sum_{q}\mathcal{F}_3^{q,a_z}(s)\,\tilde{C}^{q}_{ew}\!\left[\bar{w}_{2}\, H_{1}\, \bar{H}_{1T}^{\perp}\right],\\[0.5cm]
&F_{TT}^{\sin(2\phi+\phi_{a}-\phi_{b}),ew}  =  \sum_{q}\mathcal{F}_3^{q,a_z}(s)\,\tilde{C}^{q}_{ew}\!\left[w_{2}\, H_{1T}^{\perp}\, \bar{H}_{1}\right],
\hspace{0.35cm}F_{TT}^{\sin(2\phi+\phi_{a}+\phi_{b}),ew}  =  \sum_{q}\mathcal{F}_3^{q,a_z}(s)\,\tilde{C}\!\left[w_{4}^\prime\, H_{1T}^{\perp}\, \bar{H}_{1T}^{\perp}\right].\label{e:Fewlast}\\[-0.35cm] \nonumber
\end{align}
And for the case of polarized leptons (Eq.$\,$(\ref{eq:DefPolLept})) we find that one can obtain the values of the structure functions by making the following replacements in Eqs.$\,$(\ref{e:Few1})--(\ref{e:Fewlast}).  (Note again that we use $G$ to label structure functions associated with polarized leptons.)
\begin{equation}
\begin{array}{ccc}
F &\longrightarrow & G \\
\hline\\[-0.3cm]
\mathcal{F}_1^{q\pm}(s) &\longrightarrow& \mathcal{F}_5^{q\pm}(s) \\[0.3cm]
\mathcal{F}_2^{q}(s) &\longrightarrow& \mathcal{F}_4^{q}(s)\\[0.3cm]
\mathcal{F}_3^{q,a_z}(s) &\longrightarrow& \mathcal{F}_3^{q}(s)\\[0.3cm]
\mathcal{F}_4^{q}(s) &\longrightarrow& \mathcal{F}_2^{q}(s)\\[0.3cm]
\mathcal{F}_5^{q+}(s) &\longrightarrow& \mathcal{F}_1^{q+}(s)
 \end{array} \label{e:PolRepl}
\end{equation}
Where possible, we have compared our electroweak results to \cite{Boer:2008fr, Boer:1997qn, Boer:1999mm}.  The only disagreement is that for structure functions where the antisymmetric convolution $\tilde{C}_{ew}$ shows up, Refs.$\!$\cite{Boer:2008fr, Boer:1997qn, Boer:1999mm} have the symmetric convolution $C_{ew}$.  Again we note that in \cite{Boer:2008fr, Boer:1997qn, Boer:1999mm} different conventions for the azimuthal angles are used.

\section{Electroweak twist-2 structure functions for hadron-jet and single hadron production} \label{a:hjet}
In this Appendix we give the twist-2 cross section for $e^+e^- \!\!\rightarrow h\,jet\,X$ and $e^+e^- \!\!\rightarrow h\,X$.\footnote{As before, we only explicitly write out the results for unpolarized leptons.  One can obtain the polarized lepton expressions from the prescription in (\ref{e:PolRepl}).} The former can be found from Eq.$\,$(\ref{e:angdecompEW}) by setting $\bar{D}_1(z_b,z_b^2\vec{p}_{b\perp}^{\,2}) = \delta^{(2)}(\vec{p}_{b\perp})\delta(1-z_b)$.  (Terms that do not contain $\bar{D}_1$ are zero.)  This leads to
\begin{align}
4&\frac{P_{h}^{0}P_{J}^{0}d\sigma_{ew}}{d^{3}\vec{P}_{h}\,d^{3}\vec{P}_{J}} = zN_c\,\delta(1-z_J)\bigg\{\!\left[(1+\cos^2\theta)F_{UJ}^{1,ew} + (\cos\theta)F_{UJ}^{2,ew}\right] \nonumber\\
&\hspace{4.75cm}+ \Lambda_h\left[(1+\cos^2\theta)F_{LJ}^{1,ew} + (\cos\theta)F_{LJ}^{2,ew}\right]\nonumber\\
&\hspace{4.75cm}+ |\vec{S}_{h\perp}|\left[\sin\phi_{S_h}\left((1+\cos^2\theta)F_{TJ}^{1,ew} + (\cos\theta)F_{TJ}^{2,ew}\right)\right.\nonumber\\
&\hspace{6.25cm} \left.+\, \cos\phi_{S_h}\left((1+\cos^2\theta)\bar{F}_{TJ}^{1,ew} + (\cos\theta)\bar{F}_{TJ}^{2,ew}\right)\right]\!\bigg\}\,.\label{e:hjetX}
\end{align}
The values of the structure functions in (\ref{e:hjetX}) are given by
\begin{align}
&F_{UJ}^{1,ew} = \sum_{q,\bar{q}}\mathcal{F}_1^{q+}(s)\,D_1^{h/q}(z,z^2\vec{q}_\perp^{\,2})\,,
\hspace{0.35cm} F_{UJ}^{2,ew} = 2\widetilde{\sum_{\!q,\bar{q}}}\mathcal{F}_2^{q}(s)\,D_1^{h/q}(z,z^2\vec{q}_\perp^{\,2})\,,\\[0.5cm]
&F_{LJ}^{1,ew} = \widetilde{\sum_{\!q,\bar{q}}}\mathcal{F}_4^{q}(s)\,G_{1L}^{h/q}(z,z^2\vec{q}_\perp^{\,2})\,,
\hspace{0.35cm} F_{LJ}^{2,ew} = 2\sum_{q,\bar{q}}\mathcal{F}_5^{q+}(s)\,G_{1L}^{h/q}(z,z^2\vec{q}_\perp^{\,2})\,,\\[0.5cm]
&F_{TJ}^{1,ew} = \sum_{q,\bar{q}}\mathcal{F}_1^{q+}(s)\,\frac{q_\perp} {M_h}\,D_{1T}^{\perp\,h/q}(z,z^2\vec{q}_\perp^{\,2})\,,
\hspace{0.35cm} F_{TJ}^{2,ew} = 2\widetilde{\sum_{\!q,\bar{q}}}\mathcal{F}_2^{q}(s)\,\frac{q_\perp} {M_h}\,D_{1T}^{\perp\,h/q}(z,z^2\vec{q}_\perp^{\,2})\,,\\[0.5cm]
&\bar{F}_{TJ}^{1,ew} =\widetilde{\sum_{\!q,\bar{q}}}\mathcal{F}_4^{q}(s)\,\frac{q_\perp} {M_h}\,G_{1T}^{h/q}(z,z^2\vec{q}_\perp^{\,2})\,,
\hspace{0.35cm} \bar{F}_{TJ}^{2,ew} = 2\sum_{q,\bar{q}}\mathcal{F}_5^{q+}(s)\,\frac{q_\perp} {M_h}\,G_{1T}^{h/q}(z,z^2\vec{q}_\perp^{\,2})\,,
\end{align}
where $\widetilde{\sum\limits_{\!q,\bar{q}}}\, D^{h/q} \equiv (D^{h/q} - D^{h/\bar{q}})$, and we understand $q_\perp$ in the hadronic $cm$ frame.  

One can obtain the cross section for $e^+e^- \!\!\rightarrow \!h\,X$ by using $d^3\vec{P}_h/P_h^0 = (dz/z)(z^2d^2\vec{q}_\perp)$ (since $p_{a\perp}\equiv\vec{p}_{\perp}\! = \vec{q}_\perp$) and $d^3\vec{P}_J/P_J^0 = (q^2/4)(z_Jdz_J\,d(\cos\theta)\,d\phi)$, where $\theta$ is the scattering angle of the hadron relative to the incoming leptons (in their rest frame) and $\phi$ is its azimuthal angle.  This leads to 
\begin{align}
&\frac{d\sigma_{ew}}{dz\,d(\cos\theta)} = \frac{\pi N_c \,q^2} {8}\bigg\{\!\left[(1+\cos^2\theta)F_{U}^{1,ew} + (\cos\theta)F_{U}^{2,ew}\right] \nonumber\\
&\hspace{3.75cm}+ \Lambda_h\left[(1+\cos^2\theta)F_{L}^{1,ew} + (\cos\theta)F_{L}^{2,ew}\right]\!\bigg\},\label{e:hX}
\end{align}
where
\begin{align}
&F_{U}^{1,ew} = \sum_{q,\bar{q}}\mathcal{F}_1^{q+}(s)\,D_1^{h/q}(z)\,,
\hspace{0.35cm} F_{U}^{2,ew} = 2\widetilde{\sum_{\!q,\bar{q}}}\mathcal{F}_2^{q}(s)\,D_1^{h/q}(z)\,,\\[0.5cm]
&F_{L}^{1,ew} = \widetilde{\sum_{\!q,\bar{q}}}\mathcal{F}_4^{q}(s)\,G_{1L}^{h/q}(z)\,,
\hspace{0.35cm} F_{L}^{2,ew} = 2\sum_{q,\bar{q}}\mathcal{F}_5^{q+}(s)\,G_{1L}^{h/q}(z)\,.
\end{align}

\end{appendices}

%
%
%

\end{document}